\documentclass[lettersize,journal]{IEEEtran}
\usepackage{amsmath,amsfonts}
\usepackage{algorithmic}
\usepackage{algorithm}
\usepackage{array}
\usepackage[caption=false,font=normalsize,labelfont=sf,textfont=sf]{subfig}
\usepackage{textcomp}
\usepackage{stfloats}
\usepackage{url}
\usepackage{verbatim}
\usepackage{graphicx}
\usepackage{cite}

\usepackage{amsmath,amsfonts,bm}

\def\Maskk{\mM_{+}^{(k)}}
\def\Maskkpre{\hat{\mM}_{+}^{(k)}}
\def\Resk{\mR_{+}^{(k-1)}}
\def\Compkhat   {\hat{\mU}_{+}^{(k)}}
\def\compk{\vu^{(k)}}
\def\compkhat{\hat{\vu}^{(k)}}

\def\eqref#1{equation~\ref{#1}}

\def\1{\bm{1}}

\def\vr{{\bm{r}}}

\def\vu{{\bm{u}}}

\def\vx{{\bm{x}}}

\def\mM{{\bm{M}}}

\def\mR{{\bm{R}}}

\def\mU{{\bm{U}}}

\def\mX{{\bm{X}}}

\DeclareMathAlphabet{\mathsfit}{\encodingdefault}{\sfdefault}{m}{sl}
\SetMathAlphabet{\mathsfit}{bold}{\encodingdefault}{\sfdefault}{bx}{n}

\usepackage{cleveref}

\usepackage{algorithm}
\usepackage{algorithmic}

\usepackage{xspace}
\usepackage{booktabs}
\usepackage{enumitem}
\usepackage{multirow}
\usepackage{subcaption}
\usepackage{url}

\usepackage{xcolor}
\newcommand{\OUR}{IDSD}

\begin{document}

\title{IDSD: Iterative Deep-Learning-Based Signal Decomposition}

\author{Iris Huijben, Joël Karel, Ralf Peeters, and Pietro Bonizzi
\thanks{All authors are with the Department of Advanced Computing Sciences (DACS) at Maastricht University, The Netherlands.}
}



\maketitle

\begin{abstract}
Data-driven signal decomposition methods decompose a signal into its underlying components in a flexible and adaptive way, taking into account the signal characteristics. Here we focus on univariate signals, whose decomposition is ill-posed. Classical univariate approaches -- like variational mode decomposition -- constrain the solution space (e.g. through narrowband priors), and often require the number of components to be known in advance. These assumption, however, limit the algorithm's usage in certain real-life applications. Instead, we exploit the flexibility of neural networks to replace fixed (narrowband) priors with data-driven priors. Our model, called Iterative Deep-Learning-Based Signal
Decomposition (\OUR), iteratively extracts an adaptive number of various types of components from a signal, with no restrictions on the bandwidth of a component. We show superior performance of IDSD both in a controlled setup with synthetic data, and on two real datasets concerning tidal waves and physiological measurements.
\end{abstract}

\section{Introduction}

Analyzing real-world dynamical systems often involves interpretation of non-stationary signals. This interpretation can be facilitated by decomposing the measurement into simpler components that preserve a semantic relationship with the underlying system. Such a decomposition has many applications. For instance, in machine condition monitoring, bearing vibration signals consist of multiple oscillatory components associated with different mechanical elements~\cite{antoni2006blind}, and seismic measurements may combine a baseline signal with intermittent components related to transient activity~\cite{Bonizzi2014SingularDecomposition}. The complexity and non-stationarity of such measurements motivates the development of flexible and adaptive algorithms.

Empirical mode decomposition (EMD)~\cite{Huang1998TheAnalysis} was presented as such an adaptive method. Although promising, it suffers from mode mixing, where content of several physical components is mixed in one extracted component. Variational mode decomposition (VMD)~\cite{Dragomiretskiy2014VariationalDecomposition} and singular spectrum decomposition (SSD)~\cite{Bonizzi2014SingularDecomposition} improve on this aspect. However, SSD's performance depends on the signal's length, and VMD has an additional hyperparameter that trades off data fidelity and narrowbandness, which does not enable extracting broadband and narrowband components at the same time. Successive VMD (SVMD)~\cite{Nazari2020SuccessiveDecomposition} adaptively estimates this parameter per mode, however it still relies on a narrowband prior. Opposed to SSD and (S)VMD, our goal is to decompose a univariate signal in a set of not necessarily narrowband but, e.g., also broadband and intermittent components. Recently-proposed sparse random mode decomposition (SRMD)~\cite{Richardson2024SRMD:Decomposition} exploits sparsity of the univariate signal in time-frequency domain, and can extract different types of components using density-based clustering. While SRMD can in theory deal with broadband components, it inherently splits an intermittent component in multiple ones due to its absence at multiple moments in time. 

Advances in single-channel (monaural) speech separation have shown the promise of deep learning (DL)~\cite{Wang2018SupervisedOverview,Yang2025MultiscaleSeparation}. However, proposed methods are tailored for audio data and typically deal with maximally two or three sources. Recently, DL has also started to find its way for adaptive signal decomposition methods that do not focus on one signal modality specifically. Proposed methods are Neural Mode Estimation (NME)~\cite{Sun2023NeuralEstimation}, Iterative Residual Convolutional Neural Network (IRCNN)~\cite{Zhou2024IRCNN:Network} and its extension IRCNN$^{+}$~\cite{Zhou2025IRCNN:Decomposition}. In the approach we propose in this work, we also leverage a DL solution. The main reason to use DL for signal decomposition is the model's ability to learn priors from data, resulting in a more flexible decomposition method than classical approaches that need to set fixed (e.g. narrowband) priors. 

A downside of IRCNN~\cite{Zhou2024IRCNN:Network} and IRCNN$^{+}$~\cite{Zhou2025IRCNN:Decomposition} is that these methods rely on a neural network that directly predicts components (in time domain), making these predictions unconstrained. This can be problematic, especially for DL models, as there is no guarantee that predicted components are actually part of the input signal (i.e. they could be `hallucinated'). NME resolves this by predicting masks in the discrete cosine transform (DCT) domain using a multi-layer perceptron (MLP). Our proposed method also makes use of mask predictions, but we adopt a more structured convolutional neural network, i.e. a U-Net architecture~\cite{Ronneberger2015U-net:Segmentation}. The shift-equivariance of convolutions facilitates usage of our method on variable-length time series, while NME's MLP is restricted to a fixed number of samples. Another downside of NME and IRCNN$^{(+)}$ is their assumption that a fixed (and same) number of $K$ components is present in the signals both during training and deployment~\cite{Sun2023NeuralEstimation,Zhou2024IRCNN:Network,Zhou2025IRCNN:Decomposition}. This requires re-training the model whenever $K$ changes, and it makes the number of trainable parameters dependent on $K$. Our method improves on this aspect, as it extracts an adaptive data-driven number of test-time components using one single trained model. 

NME's mask prediction in DCT can be seen as a subset selection method. Another line of methods also constrains the solution space of extracted components by imposing sparsity in a given basis. Sparse non-negative matrix factorization finds a basis that represents the data under a sparsity constraint on component activations~\cite{LeRoux2015SparseDone}, while the non-linear matching pursuit algorithm operates on an over-complete dictionary to sparsely select atoms~\cite{Hou2013Data-drivenAnalysis}. Morphological component analysis similarly promotes sparsity using heuristically chosen dictionaries for different components~\cite{Starck2005ImageApproach}. In contrast, the empirical wavelet transform (EWT) constructs a data-adaptive filter bank by partitioning the Fourier spectrum according to heuristically determined frequency boundaries, yielding components supported on adaptively selected frequency bands~\cite{Gilles2013EmpiricalTransform}. A common limitation of these approaches is that their decomposition depends on either a predefined basis or dictionary, or data-dependent design choices that may be application-specific. In contrast, our method leverages the complete discrete Fourier basis, which provides an exact and invertible representation of any finite-length discrete signal.

In this work, we present Iterative Deep-learning-based Signal Decomposition (\OUR) and show its superior performance both in a controlled setup and on real data. It has the following properties: 1) Thanks to IDSD's DL nature, it can learn flexible component priors from a training set of data, therewith supporting different type of components (narrowband, broadband, intermittent, among others). 2) In the absence of signals with their ground-truth components, IDSD can be trained on synthetic signals only. 3) \OUR~predicts, for each component, a mask in frequency domain. A component's spectrum is, therefore, constrained to be a subset of the signal's spectrum. 4) Thanks to using the complete Fourier dictionary, \OUR~does not rely on modality- or application-specific assumptions for choosing a transform domain or dictionary. 5) During deployment, \OUR~iteratively extracts an adaptive number of components $K$ (until a stopping criterion is met), which does not need to be known during training. It, therefore, only requires training one model regardless of $K$, with the model size being independent of $K$.

\begin{figure*}
    \centering
    \includegraphics[page=8,width=1\linewidth,trim={0cm 15cm 7.9cm 0cm},clip]{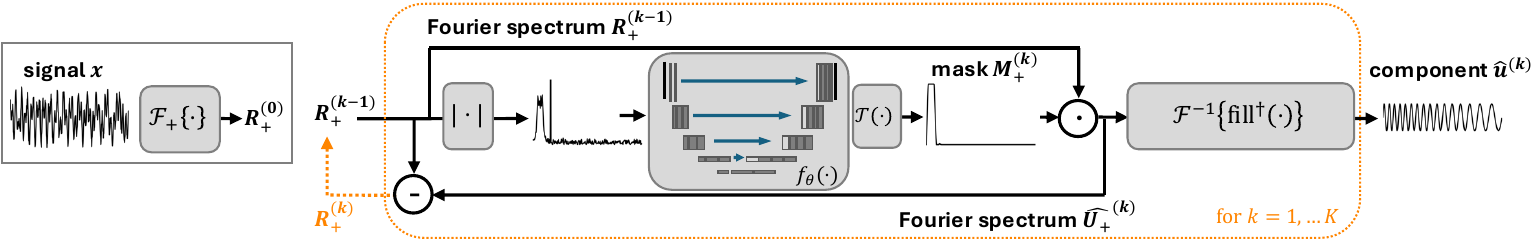}
    \caption{Signal decomposition pipeline of IDSD during deployment: the trained $f_{\theta}$ (a U-Net architecture) iteratively predicts masks in frequency domain that pass through the frequency bins belonging to each component.}
    \label{fig:pipeline}
\end{figure*}

\section{Problem definition}
We define a univariate signal $\vx \in \mathbb{R}^N$, with $N$ time steps sampled at $f_s$~Hz for $T=N/f_s$ s. It contains \mbox{$K\!+\!1$} components: $\vx\!=\!\sum_{k=1}^{K+1}\vu^{(k)}$, with components $1,\ldots,K$ related to the underlying system, and the last component $K+1$ related noise/residual component. The components are of the form:

\begin{equation}
u^{(k)}[n] = a[n] \cos(\phi[n]),
\end{equation}
with an instantaneous amplitude $a[n]$ and an instantaneous phase $\phi[n]$. In the following, we will use the terms `components' and `modes' interchangeably, as both are commonly used in literature.

Our goal is to extract underlying components $\{\vu^{(1)},\ldots,\vu^{(K)}\}$ from $\vx$ (and thus also $\vu^{(K+1)}$). This can be posed as the following optimization problem:

\begin{equation}
\label{eq:optimization}
    \hat{\vu}^{(1)},\ldots,\hat{\vu}^{(K+1)} = \underset{\vu^{(1)},\ldots,\vu^{(K+1)}}{\operatorname{argmin}} ||\vx - \sum_{k=1}^{K+1}\vu^{(k)}||.
\end{equation}
The optimal solution $\hat{\vu}^{(k)} = \vu^{(k)}$, for $k=1,\ldots,K+1$ cannot be found analytically as \cref{eq:optimization} is an under-determined system.

\section{Notation}

We define $\mX := \mathcal{F}\{\vx\} \in \mathbb{C}^{\tilde{N}}$ as a spectral representation of $\vx$ with $\tilde{N}$ frequency bins, and $|\mX| \in \mathbb{R}^{\tilde{N}}$ as its magnitude. We use subscript $+$ to denote the positive frequency axis: $\mX_{+} := \mathcal{F}_{+}\{\vx\} \in \mathbb{C}^{F}$, where $F$ is the number of frequency bins in the positive frequency axis\footnote{$F = \tilde{N}/2 + 1$ if $\tilde{N}$ is even, and $F = (\tilde{N}+1)/2$ if $\tilde{N}$ is odd.}. We use $\mathcal{U}_{[a,b]}$ to denote a uniform random variable on the interval $a$ to $b$ (including). To indicate a categorical random variable with equal probability for selecting any integer between $a$ and $b$ (including), we use $\mathcal{U}_{\{a,\dots,b\}}$.

\section{The IDSD algorithm}

\subsection{General definition}
\label{sec:general_idea}

The goal of Iterative Deep-learning-Based Signal Decomposition (IDSD) is to extract components $\vu^{(k)}$ for \mbox{$k\!=\!1,\ldots,K+1$} from signal $\vx$. In the forward signal model, we assume that $\vu^{(1)},\ldots,\vu^{(K+1)}$ originate from a generative joint distribution 
$p(\vu^{(1)},\ldots,\vu^{(K+1)})$, and that $\vx$ is deterministically given by their sum, 
inducing a joint distribution $p(\vx,\vu^{(1)},\ldots,\vu^{(K+1)})$. Inverting the forward model, one could infer $\vu^{(1)},\ldots,\vu^{(K+1)}$ from $\vx$ using the conditional distribution $p(\vu^{(1:K+1)} \mid \vx)$. We approximate this conditional via a sequential factorization:
\begin{align}
\label{eq:r_to_u}
p(\vu^{(1:K+1)} \mid \vx) \approx \prod_{k=1}^{K+1} p\big(\vu^{(k)} \mid \vr^{(k-1)}\big), \quad \nonumber \\ 
\text{with} \quad  \vr^{(0)} = \vx \quad \text{and} \quad 
\vr^{(k)} = \vr^{(k-1)} - \vu^{(k)}.
\end{align}

\noindent Modeling $p\big(\vu^{(k)} \mid \vr^{(k-1)}\big)$ directly is intractable due to its high dimensionality. In practice we approximate each $\vu^{(k)}$ by a point estimate $\compkhat$, which is a high-likelihood point under $p\big(\vu^{(k)} \mid \vr^{(k-1)}\big)$  (see \cref{sec:training_phase}). Specifically, for steps $k=1,\dots,K$, the estimated component equals:
\begin{align}
\compkhat = h_\theta(\vr^{(k-1)}),
\end{align}
where $h_\theta$ is a composition of deterministic functions, visualized in \cref{fig:pipeline}, and explained (in order) below.

First, a discrete Fourier transform is applied, and only the positive side of the spectrum is retained:

\begin{equation}
\label{eq:step1}
\Resk = \mathcal{F}_{+}\{\vr^{(k-1)}\}\in \mathbb{C}^F.
\end{equation}

Second, the magnitude is taken: $\lvert \Resk \rvert$. Third, a neural network $f_{\theta}$ with trainable parameters $\theta$ takes this magnitude and predicts a `soft' mask (or frequency domain filter):

\begin{equation}
\label{eq:step2}
\Maskkpre = f_{\theta}(\lvert\Resk\rvert ) \in [0,1]^F.
\end{equation}

Fourth, a mask post-processing function $\mathcal{T}(\Maskkpre) \in [0,1]^F$ is applied (see \cref{sec:post_process_mask}). Fifth, the post-processed mask is element-wise applied (denoted with $\odot$) on the positive-sided complex residual spectrum: 

\begin{equation}
\label{eq:step5}
\Compkhat = \mathcal{T}(\Maskkpre) \odot \Resk \in \mathbb{C}^F.
\end{equation}

Sixth, the negative frequency axis is filled with the Hermitian conjugate of the post-processed spectrum:

\begin{equation}
\label{eq:step6}
\hat{\mU}^{(k)} = \operatorname{fill}^{\dag} \big( \Compkhat \big) \in \mathbb{C}^{\tilde{N}}.
\end{equation}

Lastly, the resulting double-sided spectrum is converted to time domain to get the $k^{\text{th}}$ component:

\begin{equation}
\label{eq:step7}
    \compkhat = \mathcal{F}^{-1} \big( \hat{\mU}^{(k)} \big). 
\end{equation}

The next residual spectrum is then computed through:
\begin{equation}
\mR_{+}^{(k)} = \Resk - \Compkhat,
\end{equation}
and the aforementioned steps are repeated from \cref{eq:step2} onwards.
Extracting the last component (i.e. the noise component) is simply achieved by transforming the last residual's spectrum back to time domain: $\hat{\vu}^{(K+1)} = \mathcal{F}^{-1}\{\operatorname{fill}^{\dag}( \mR_{+}^{(K)} )\}$. As a result, the decomposition is exact, i.e., $\vx = \sum_{k=1}^{K+1} \compkhat$ is guaranteed. Moreover, thanks to the element-wise masking procedure, each component's spectrum is constrained to be a subset of the signal's spectrum. \Cref{pseudocode} provides pseudocode of the described algorithm, and a Python implementation can be found at \url{https://github.com/IamHuijben/IDSD}.
\begin{algorithm}[t]
\caption{Deployment of \OUR}
\label{pseudocode}
\begin{algorithmic}[1]  
\REQUIRE signal $\vx$
\ENSURE Signal components $[\compkhat$, \text{for}~$k\!=\!1,\ldots,K]$, and noise/residual component $\hat{\vu}^{(K+1)}$
    \STATE $k=0$
    \STATE $\mX_{+} = \mathcal{F}_{+}\{\vx\}$
    \STATE $\mR_{+}^{(0)} = \mX_{+}$  \textit{\small{\hfill// First residual spectrum is signal spectrum}} 
    \WHILE{not $\operatorname{stopping}$}  
        \STATE $k = k + 1$
        \STATE $\Maskkpre = f_{\theta}(|\Resk|)$
        \STATE $\Compkhat = \mathcal{T}(\Maskkpre) \odot \mR_{+}^{(k-1)}$   
        \STATE $\mR_{+}^{(k)} = \Resk - \Compkhat$
        \STATE check stopping criterion
    \ENDWHILE
    \STATE $K = k$
    \STATE $\compkhat = \mathcal{F}^{-1}\{\operatorname{fill}^{\dag}(\Compkhat)\}$ for $k\!=\!1,\ldots,K$
    \STATE $\hat{\vu}^{(K+1)} = \mathcal{F}^{-1}\{\operatorname{fill}^{\dag}(\mR_{+}^{(K)}\}$
\end{algorithmic}
\end{algorithm}
\subsection{Architecture of neural network $f_\theta$}
\label{sec:architecture}
\begin{figure}
    \centering
    \includegraphics[page=10,width=\linewidth,trim={0cm 4cm 5cm 0cm},clip]{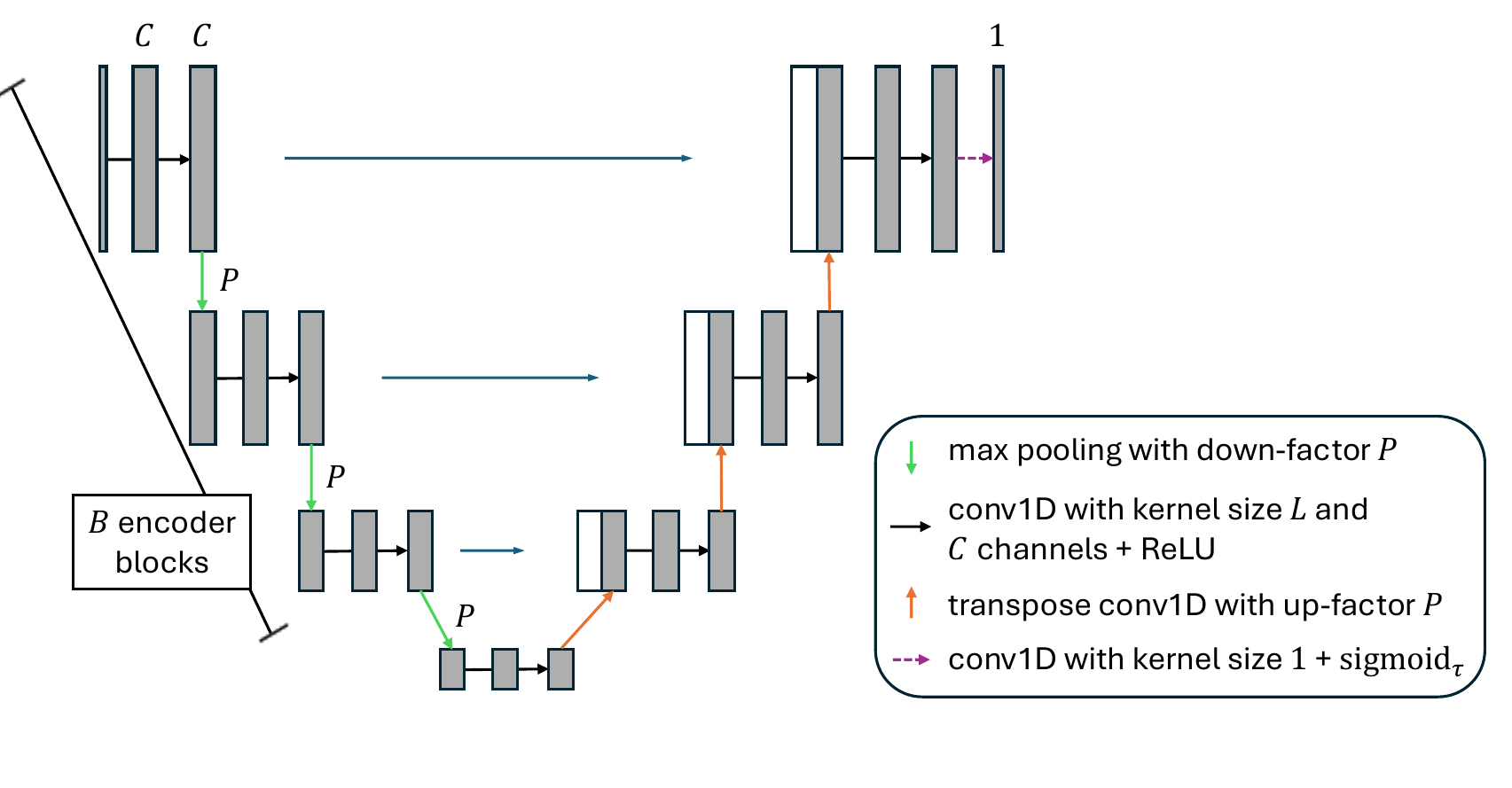}
    \caption{U-Net structure as used for $f_\theta$, with the $C$ convolutional channels per convolutional layer, with kernels of length $L$, $B$ number of convolutional blocks in encoder/decoder, and pooling factor and up-sampling factor $P$.}
    \label{fig:Unet}
\end{figure}
Neural network $f_{\theta}$ transforms a spectrum's magnitude to a soft mask (i.e. with values between 0 and 1) of the same dimensionality: $\Maskkpre = f_\theta(|\Resk|)$. This is a classic segmentation task, for which the U-Net architecture~\cite{Ronneberger2015U-net:Segmentation} is a popular choice. U-Net is an encoder-decoder style neural network, where the encoder progressively down-samples the input, and the decoder again up-samples the latent representation to an output of the same dimensionality as the input. Skip connections link corresponding encoder and decoder layers, allowing the network to combine both coarse information and high-resolution details.

The original U-Net architecture has fixed settings and uses 2D convolutional kernels~\cite{Ronneberger2015U-net:Segmentation}. We use 1D kernels instead (as our magnitude spectrum is 1D), and we parameterize some of the settings (see \cref{fig:Unet}): $B$ is the number of encoding (and decoding) blocks, where each encoder block contains two ReLU-activated convolutional layers with a 1D kernel of length $L$. Each encoder block ends with a max-pooling layer with pooling factor $P$ (implemented as: stride=$P$ and kernel size=$P$). Each decoder block contains a transpose 1D convolution that up-samples its input with a factor $P$, followed by two ReLU-activated 1D convolutions with kernels of length $L$. A bottleneck block is present between the encoder and decoder, being similar to an encoder block, except for not having a pooling layer. After the last decoder block, a last 1D convolutional layer follows (again with kernel size $L$), activated by a $\operatorname{sigmoid}_{\tau}$ function with temperature parameter $\tau$. Biases were turned off in the entire U-Net to minimize the ability to learn a bias towards specific frequency bins. 

Setting $C$, $B$, $P$ and $L$ affects the number of trainable parameters, and the effective receptive field of the model. The effective receptive field is defined as the region of the input that can influence an output element through the network’s sequence of convolutional and down-sampling operations~\cite{Le2018WhatNetworks}. We set $(B,P,L)=(3,3,7)$ as it results in an effective receptive field that covers the entire input spectrum (see Appendix \ref{app:receptive_field}), which was shown to be beneficial for U-Net architectures~\cite{Behboodi2020ReceptiveU-Net}. Furthermore, unless stated otherwise, we report performance for the model trained with $C=128$.
\begin{table*}
\footnotesize
\centering
\caption{Different type of synthetic components with their amplitude and frequency definitions. Settings were: $a_{\text{min}} = 0.1$, $a_{\text{max}} = 1.0$, $f_{\text{min}} = 1$~Hz, $f_{\text{max}} = f_s/2-1$~Hz, $f_s=1024$~Hz, $N=1024$ samples.}
\begin{tabular}{p{2.8cm}||p{8.9cm}|p{4.3cm}}
 & \bf Amplitude $a_k[n]$ & \bf Frequency $f_k[n]$ \\
\midrule
 \bf A) Sinusoid  &   $\mathcal{U}_{[a_{\text{min}},a_{\text{max}}]}$\vspace{0.05cm}& $\mathcal{U}_{[f_{\text{min}},f_{\text{max}}]}$\vspace{0.05cm}  \\
\hline
 \vspace{0.03cm}
\bf  B) AM component &   $a_{\text{base}}\big(a_{\text{start}} + \delta a \cdot n  \big)$, \hspace{0.3cm} with \hspace{0.3cm} $a_{\text{base}}\!\sim\!\mathcal{U}_{[a_{\text{min}},a_{\text{max}}]}$, 
 
 $\delta a = (a_{\text{end}}\!-\! a_{\text{start}})/N$, $a_{\text{start}}\sim\mathcal{U}_{[0.1,1.0]}, a_{\text{end}}\sim\mathcal{U}_{[0.1,1.0]}$ & 
  \vspace{0.03cm}$\mathcal{U}_{[f_{\text{min}},f_{\text{max}}]}$ \\
\hline
 \vspace{0.03cm}
\bf  C) FM component & 
 \vspace{0.03cm}
 $\mathcal{U}_{[a_{\text{min}},a_{\text{max}}]}$ &  $\operatorname{clip}\big(f_{\text{base}} + \delta f \cdot n$, $f_{\text{min}}$, $f_{\text{max}}$\big), with 
 
 $\delta f \sim \mathcal{U}_{[-0.1f_{\text{max}},0.1f_{\text{max}}]} /N$ \\
\hline
\textbf{D) Intermittent component}

\vspace{0.3cm}
D1) Gaussian bursts 

\vspace{0.4cm}
D2) On/off bursts& 
$a_{\text{base}} \cdot \operatorname{clip}\big(\sum_{b=1}^{B} g(\mu_b,\sigma_b)[n], 0, 1 \big)$, with 

$a_{\text{base}} \sim \mathcal{U}_{[a_{\text{min}},a_{\text{max}}]}$, $B \sim \mathcal{U}_{\{1,\ldots,5\}}$, $\mu_b \sim \mathcal{U}_{[0,N]}$, $\sigma_b \sim \mathcal{U}_{[0.01N,0.1N]}$,

$g(\mu_b,\sigma_b)[n] = \frac{\exp\big(-(n-\mu_b)^2\big)}{2\sigma_b^2}$

$ g(\mu_b,\sigma_b)[n] =
\begin{cases}
  1 & \text{for } n= \big[\mu_b-\lfloor \frac{\sigma_b}{2} \rfloor,\ldots,\mu_b+ \lfloor \frac{\sigma_b}{2} \rfloor \big] , \\
  0 & \text{otherwise}
\end{cases}$ 
\vspace{-0.07cm} 
& \vspace{0.7cm}$\mathcal{U}_{[f_{\text{min}},f_{\text{max}}]}$ 

\vspace{0.4cm}$\mathcal{U}_{[f_{\text{min}},f_{\text{max}}]}$ \vspace{-0.07cm} 
\\
\bottomrule
\end{tabular} 
\label{tab:data_generator}
\end{table*}

\subsection{Training phase of neural network $f_\theta$}
\label{sec:training_phase}

\subsubsection{Training data}

The trainable parameters in $\theta$ require to be tuned, which is done through a one-time training phase. The assumption we make is that the joint distribution $p_{\text{train}}(\vx,\vu^{(1)},\ldots,\vu^{(K+1)})$ (hereafter~$p_{\text{train}}$) from which training data are generated, is a good approximation of the distribution $p_{\text{test}}$ from which the test-time signals originate. To cover different type of signals, we generate training signals as a summation of a broad set of narrowband, broadband, non-stationary, and intermittent components. To this end, we adopt the following generative signal model:
\begin{equation}    
\label{eq:signal_model}
x[n] = \sum_{k=1}^{K} a^{(k)}[n]\cos\Big(2\pi\frac{f^{(k)}[n]}{f_s}n + \phi^{(k)} \Big) + \mathcal{N}(0,\sigma),
\end{equation}
\noindent with $f_s=1024$~Hz, $T=1$~s, $\phi^{(k)} \sim  \mathcal{U}_{[0,2\pi]}$, and $\mathcal{N}(0,\sigma)$ being additive white Gaussian noise (AWGN) with $\sigma \sim \mathcal{U}_{[0,0.2]}$. Using this signal model, we create different component types by varying $a^{(k)}[n]$ and $f^{(k)}[n]$, resulting in either: A) a pure sinusoid, B) an amplitude-modulated (AM) component, C) a frequency-modulated (FM) component, or D) an intermittent component, either with Gaussian bursts of activity (D1) or on-off bursts (D2), see \cref{tab:data_generator}.

During training, we generate data on the fly, with a varying number of components per signal by sampling $K\sim\mathcal{U}_{\{1,\ldots,5\}}$ for each generated signal. The signal always contains one broadband FM component (C). If $K>1$, another component is uniformly selected from types A, B or D. In case of an intermittent component (D), it is randomly selected to be of type D1 or D2. All further $K\!-\!2$ components, if present, are pure sinusoids (A).

\subsubsection{Optimization}
\label{sec:loss_function}

As the training data are synthetically generated data, the components that are part of each signal are known. We can, therefore, train the model (i.e. update its parameters $\theta$) in a supervised fashion. To this end we minimize the mean-squared-error (MSE) between the dominant ground-truth component $\mU_{+}^{(1)}$ and the first extracted component $\hat{\mU}_{+}^{(1)} = \mM_{+}^{(1)} \odot \mX_{+}$. The dominant ground-truth component is defined as the one with the highest peak in frequency domain\footnote{Note that the dominant ground-truth component can be any component in the signal as this is based on the height of the peaks in frequency domain. Selecting the dominant component is, therefore, unrelated to the order in which components are generated by the generator.}. The loss ($\propto$ MSE) is defined as: 

\begin{equation}
\label{eq:loss}
\mathcal{L} = \mathbb{E}_{p_{\text{train}}} \Big[\|\mU_{+}^{(1)} - \mM_{+}^{(1)} \odot \mX_{+}\|^2_2 \Big].
\end{equation}

\noindent Minimizing this loss, corresponds to maximizing the likelihood of $\mU_{+}^{(1)}$ given $\mX_{+}$ under a Gaussian error assumption.

Note that instead of only using the dominant ground-truth component $\mU_{+}^{(1)}$ in the training objective, the model could also be trained on all components, by sequentially extracting $K$ components from the signal (in the same way as done during deployment, see \cref{sec:general_idea}), and penalizing the error between the ground-truth components and the extracted ones. In that way the likelihood of $\mU_{+}^{(k)}$ given $\Resk$ would be maximized for all $k$ (in $1,\ldots,K$). Yet, we assume that likelihood maximization of only $\mU_{+}^{(1)}$ given $\mR_{+}^{(0)} (=\mX_{+})$ is sufficient to learn a mapping that can also predict high-likelihood components for $k > 1$. In \cref{sec:ablation_multiple_losses} we test our assumption that only using the dominant component for training is sufficient.

We empirically approximate the expectation over the training distribution in \cref{eq:loss} by using batches of 256 sampled (i.e. generated) training signals (and their corresponding ground-truth components). Optimization was done using the Adam optimizer \cite{Kingma2015Adam:Optimization}. We set a learning rate of $\mbox{1e-4}$, and reduced it a factor 10 when the validation loss plateaud. The validation loss is computed on randomly-generated batches of validation signals, which are generated using the same generator as the one that generated training data. The model weights at the epoch with the lowest validation loss were used to report performance on separate test sets/signals (see \cref{sec:test_signals}). The full pipeline was run with three different seeds for randomization. The seed influences the initialization of the trainable parameters in $\theta$, and the data generator used during training. We report performance of these three runs separately to analyze the variance in performance caused by seeding.

\subsection{Input spectrum pre-processing}
\label{sec:representationX}

\OUR~takes the positive side of a spectral representation of $\vx$ as input. We here describe the steps used to acquire this representation. First, the Discrete Fourier Transform (DFT) is applied. As the chosen U-Net architecture of $f_\theta$ induces a total down-sampling factor of $P^L$ (see \cref{sec:architecture}), the total number of frequency bins in the positive side of the spectrum should be integer-divisible by $P^L$. As such, we use an $\tilde{N}$-point DFT, with $\tilde{N}$ the nearest integer larger or equal to $N$, such that $F$ -- the number of bins in half the spectrum -- is integer divisible by $P^L$. Lastly, the magnitude of the positive-sided spectrum is normalized by its standard deviation. 

\begin{figure*}
    \centering
    \includegraphics[page=9,width=1\linewidth,trim={0cm 12.5cm 0cm 0cm},clip]{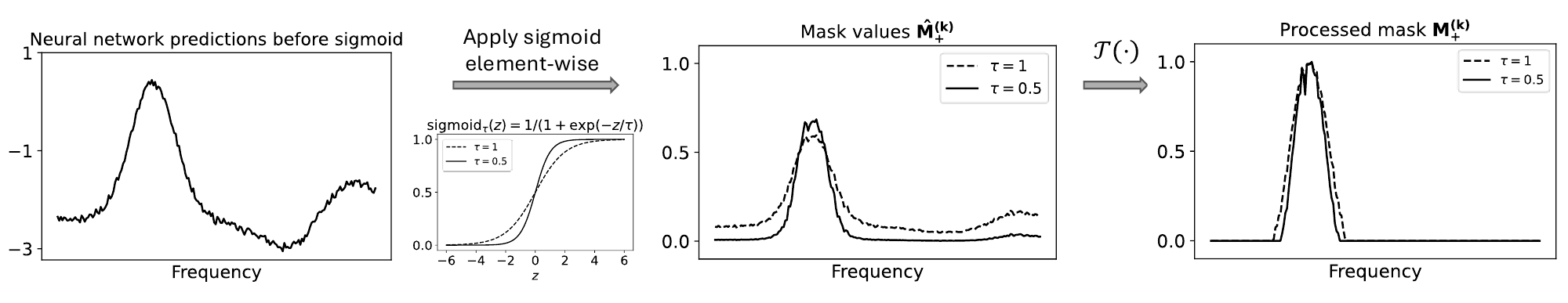}
    \caption{An illustrative example of the neural network's output before its final sigmoid activation, in which values for all frequency bins are unbounded (left). The sigmoid activation bounds them between 0 and 1 (middle). Dependent on temperature $\tau$, the mask boundaries get steeper (for lower $\tau$) or less steep (for higher $\tau$). Post-processing function $\mathcal{T}$ ensures a minimum and maximum value of 0 and 1, respectively, and one contiguous region such that a single component is extracted per step (right).}
    \label{fig:post_process}
\end{figure*}

\subsection{Mask post-processing}
\label{sec:post_process_mask}

The soft mask $\Maskkpre = f_{\theta}(|\mR_{+}^{(k)}|) \in [0,1]^F$ has values between 0 and 1 thanks to the final sigmoid activation function used in neural network $f_{\theta}$. The tempered-sigmoid function $g_{\tau}(z) = \frac{1}{1+\exp(-z/\tau)}$ is a monotonically increasing function that maps a value $z \in \mathbb{R}$ between 0 and 1. In a neural network context it is applied element-wise on all output values of its preceding layer. Temperature parameter $\tau > 0$ is typically set to 1. However, a lower value pushes $z$ closer to 0 and 1. As a result, when setting $\tau < 1$, the predicted mask $\Maskkpre$ gets steeper boundaries (illustrated in \cref{fig:post_process}), while a higher value better facilitates overlapping boundaries between nearby masks. We train with $\tau=1$, but observed slightly better deployment performance with $\tau < 1$, so by set a default $\tau=0.5$ during deployment. 

We apply some additional post-processing on the predicted mask to get $\Maskk = \mathcal{T}(\Maskkpre)$, that filters the residual spectrum. First, the minimum value of $\Maskkpre$ is subtracted from all values in $\Maskkpre$. Second, in case the mask contains more than one non-zero contiguous region (being a number of adjacent frequency bins with non-zero mask values), only the region that covers the dominant peak in the residual spectrum is retained. 
The other non-zero regions are set to zero. Third, the mask is divided by its maximum value. As a result, the final mask $\Maskk$ at step $k$, contains a single region of adjacent frequency bins with values between 0 and 1. \Cref{fig:post_process} visually shows the result of applying this post-processing transformation $\mathcal{T}$. 

\subsection{Stopping criterion} 
\label{sec:stopping}

\OUR~uses an adaptive stopping criterion. We base this criterion on monitoring the remaining energy in the Fourier domain of the residual spectrum. Once the residual's energy drops below a given percentage $\zeta$ of the original signal's energy, mode extraction stops and all remaining energy is captured in the last residual component. We set the default value $\zeta=1\%$, as also used in SSD~\cite{Bonizzi2014SingularDecomposition}. 

In case the residual energy remains similar between two steps (i.e. $\sum_{j=0}^{F-1}|\mR_{+}^{(k)}[j]|^2 \approx \sum_{j=0}^{F-1}|\mR_{+}^{(k-1)}[j]|^2$), and the remaining energy is still above the threshold ($\sum_{j=0}^{F-1}|\mR_{+}^{(k)}[j]|^2 > \frac{\zeta}{100}\sum_{j=0}^{F-1}|\mX_{+}[j]|^2$), this adaptive criterion will not stop the algorithm, and the model would be extracting solely noise or zero-components. As such, a second stopping criterion is added, where the energy of the updated residual should decrease more than a tolerance percentage $\kappa$ of the previous residual, otherwise mode extraction is stopped as well. We set again default $\kappa=1\%$,

The fact that \OUR~can use the residual energy for an adaptive stopping criterion is thanks to the guarantee that the residual energy can not increase after extracting an additional component, which we show below. Given \cref{eq:step5}, the residual spectrum at step $k$ is defined as: 

\begin{equation}
    \mR_{+}^{(k)} = \Resk - \Compkhat = \big (\mathbf{1} - \mathcal{T}(\Maskkpre) \big) \odot \Resk .
\end{equation}

\noindent Thanks to $\mathcal{T}(\Maskkpre)$ containing values between 0 and 1, the energy of the residual spectrum cannot increase from extracting a new component, i.e.:

\begin{equation}
\label{eq:convergence}
\sum_{j=0}^{F-1} |\mR_{+}^{(k)}[j]|^2 \leq  \sum_{j=0}^{F-1} |\mR_{+}^{(k-1)}[j]|^2.
\end{equation}

\noindent Using Parseval's theorem \cite{stein2011fourier}:

\begin{equation}
\label{eq:parseval}
\frac{1}{\tilde{N}} \sum_{j=0}^{\tilde{N}-1} |\mR^{(k)}[j]|^2 = \sum_{n=0}^{\tilde{N}-1} |\vr^{(k)}[n]|^2,
\end{equation}

\noindent the symmetry of the Fourier spectrum of a real signal, and \cref{eq:convergence}, it can be seen that the time-domain energy of the residual cannot increase after extracting an additional component:

\begin{equation}
\sum_{n=0}^{\tilde{N}-1} |\vr^{(k)}[n]|^2 \leq \sum_{n=0}^{\tilde{N}-1} |\vr^{(k-1)}[n]|^2.
\end{equation}

For some use cases, the user might know the number of components $K$ in advance. If $K$ is lower than the number of extracted components, the first $K$ components can simply be taken, as \OUR's~mode extraction does not depend on setting a value for $K$. This in contrast to other methods like EWT and VMD, for which the first few extracted components may change if $K$ is set to a different value. 

\section{Experiments on simulated data}

\subsection{Test signals}
\label{sec:test_signals}

We generate test signals according to the signal model provided in \cref{eq:signal_model}, and with the same rules of generating type A-D signals as presented in \cref{sec:training_phase}. We generate two test sets of 10,000 signals each, with $K\!=\!2$, resp., $K\!=\!5$ components and AWGN with $\sigma=0.1$. 

To gain a deeper understanding of the performance for the different types of signals, we also generate three additional sets, each of 10,000 signals with $K\!=\!2$ components. Set AB contains signals that sum a pure sinusoid (type A, see \cref{tab:data_generator}), an amplitude-modulated sinusoid (type B) and AWGN ($\sigma=0.1$). Set AC includes signals containing a sinusoid, a frequency-modulated sinusoid (type C), and AWGN ($\sigma=0.1$). Lastly, set AD contains signals combining a sinusoid, an intermittent component (type D), and AWGN ($\sigma=0.1$).

\begin{figure*}
    \centering
    \includegraphics[width=1\linewidth,trim={0cm, 0cm, 0cm, 0.cm},clip]{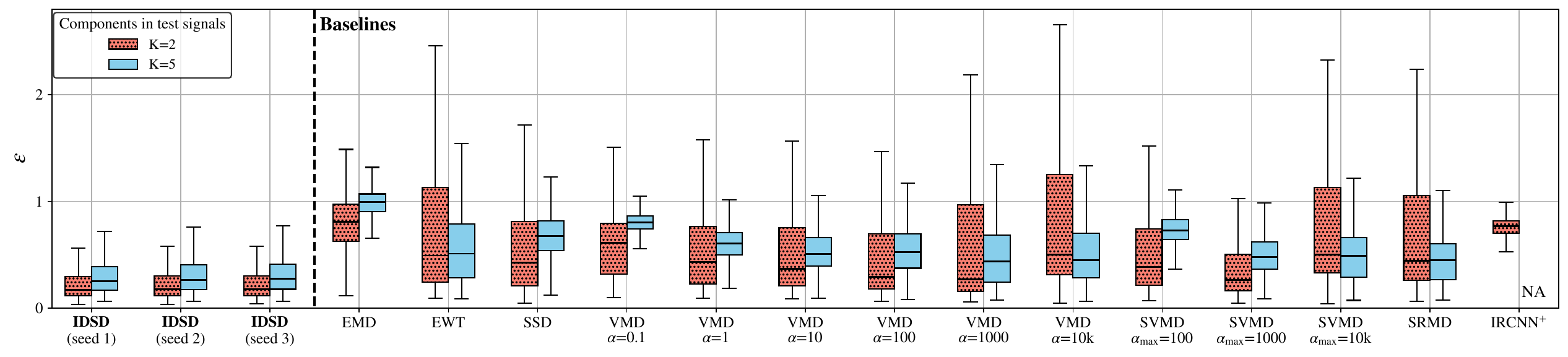}
    \caption{Normalized $\ell_2$ errors $\epsilon$ for different models on the test sets with signals containing $K\!=\!2$ and $K\!=\!5$ components. \OUR~outperforms all other algorithms, performs very similar on both test sets (while the model was not re-trained for the different number of components in each test set), and it is not sensitive to its initialization (similar performance across seeds).}
    \label{fig:synthetic_K2_K5}
\end{figure*}

\subsection{Quantitative analyses}

The use of simulated composite signals allows for an objective comparison between the extracted components and the ground-truth components used to generate the signal. To this end, we evaluate the average normalized $\ell_2$ error between the $K$ ground-truth components $\vu^{(k)}$ and their predictions $\compkhat$:
\begin{equation}
\label{eq:metric}
\epsilon = \frac{1}{K}\sum_{k=1}^{K} \epsilon^{(k)}, \text{~~~with~~~} \epsilon^{(k)} = \frac{\lVert\compk - \compkhat\rVert_2}{\lVert\compk\rVert_2}.
\end{equation}

This relative error $\epsilon$ is computed for all decomposed signals in each test set. We compare performance of \OUR~to a large variety of baselines: EMD~\cite{Huang1998TheAnalysis}, EWT~\cite{Gilles2013EmpiricalTransform} SSD~\cite{Bonizzi2014SingularDecomposition}, VMD~\cite{Dragomiretskiy2014VariationalDecomposition}, SVMD~\cite{Nazari2020SuccessiveDecomposition}, SRMD~\cite{Richardson2024SRMD:Decomposition} and $\text{IRCNN}^{+}$~\cite{Zhou2025IRCNN:Decomposition}\footnote{We use built-in Matlab implementations for EMD, EWT and VMD and the provided Matlab code for SSD and SVMD. For SRMD we used the provided Python implementation. Both SSD and SVMD determine the number of modes themselves using a stopping criterion, but to facilitate comparison, we fixed all algorithms to extract $K=2$ and $K=5$ components.}. For all algorithms, per signal a linear sum assignment problem is solved to match each predicted component to a (unique) ground-truth component, such that the total relative error is minimized. We run VMD with data-fidelity parameter $\alpha \in \{0.1,1,10,100,1000,10k\}$, and SVMD for $\alpha_{\text{max}} \in \{100,1000,10k\}$. For both algorithms, the time step of the dual-ascent was set to zero, the recommendation for noisy signals. We converted the available Keras code for $\text{IRCNN}^{+}$ to Pytorch and benchmarked our implementation against the original one\footnote{$\text{IRCNN}^{+}$ code was only provided for extracting two components.} (see Appendix \ref{app:benchmark_ircnn}). 

\Cref{fig:synthetic_K2_K5} shows the relative error of \OUR~and all baselines on the test sets with $K=2$, resp., $K=5$ components. \OUR~outperforms all baselines for both sets. Moreover, it can be seen that the seed for randomization does hardly affect performance. \Cref{fig:subsets} in Appendix \ref{app:subsets} compares the relative errors of \OUR for sets AB, AC and AD separately. 

\begin{figure*}
    \centering
\includegraphics[width=0.8\linewidth,trim={0cm, 0cm, 0cm, 0.cm},clip]{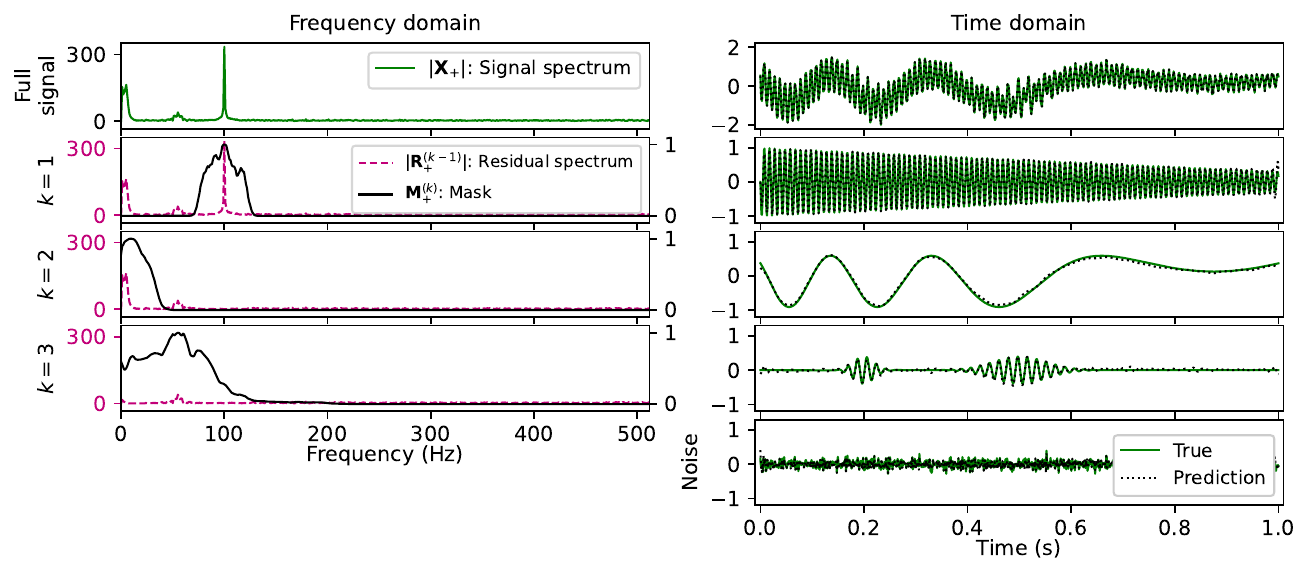}
    \caption{An example signal with an AM, FM, and intermittent component. \OUR~predicts masks (left; black) to extract the components (right; black) from the residual spectra (left; pink).}
    \label{fig:example_signal}
\end{figure*}
\subsection{Qualitative analyses}

\Cref{fig:example_signal} shows the working principle of \OUR~on an example test signal with three different types of components and noise. The left sub-figure shows at each step $k$ the residual spectrum at that step (in pink), and the predicted mask (in black) used to filter this residual. The right sub-figure shows the reconstructions in time domain after applying the mask to the residual spectra.

\begin{figure*}
    \centering
    \includegraphics[width=0.8\linewidth]{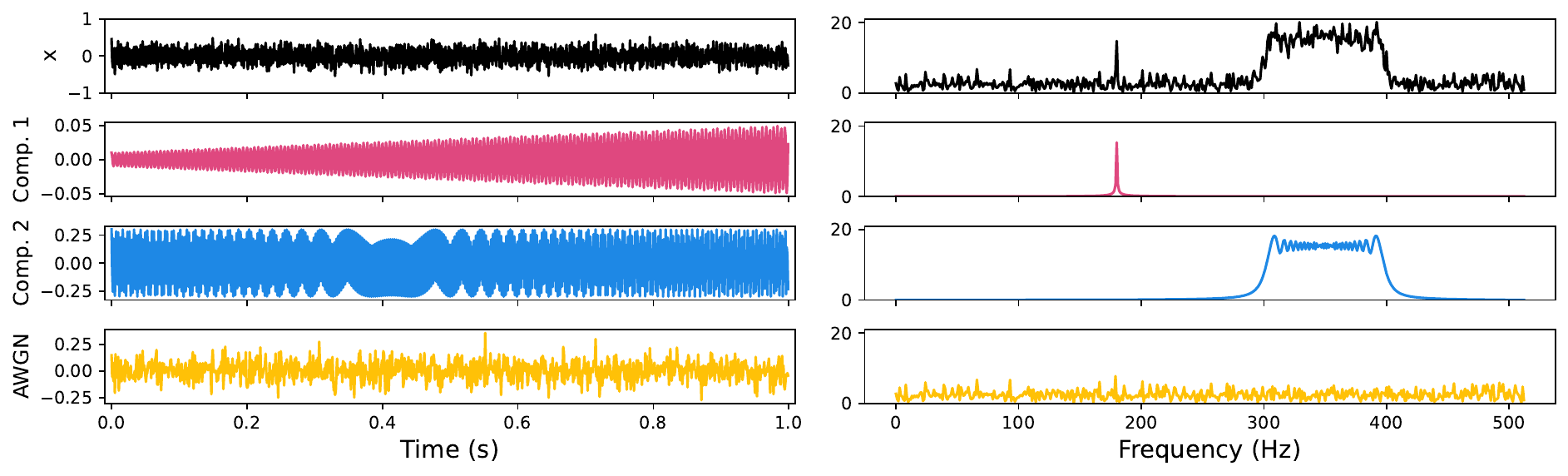}
    \caption{Signal $x_{\text{sig1}}(t)$ (top; in black) and its ground-truth components in time domain (left) and frequency domain (right).}
    \label{fig:example_BC}
\end{figure*}
\begin{figure*}
    \centering
    \includegraphics[width=\linewidth]{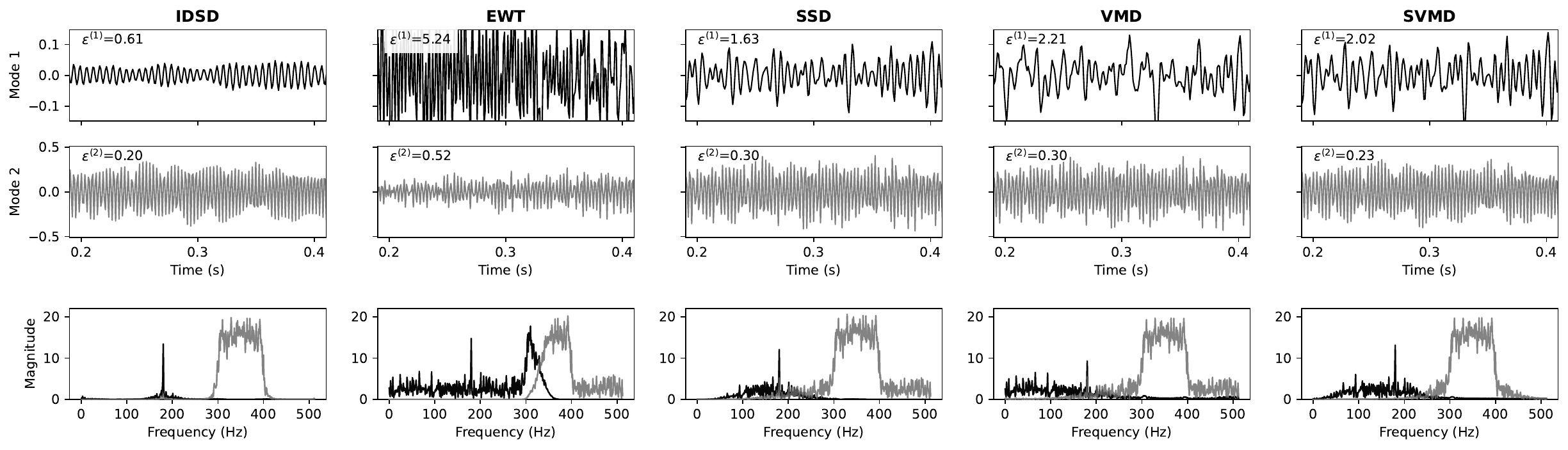}
    \caption{Decomposition of $x_{\text{sig1}}(t)$ in two components by different methods. The time axis is zoomed for better visibility.}
    \label{fig:example_BC_decomp}
\end{figure*}

We qualitative compare performance of \OUR~against EWT, SSD, VMD, and SVMD (the latter two with the best $\alpha$, resp. $\alpha_{\text{max}}$, for each given signal). We define a signal with an AM sinusoid, a FM sinusoids and AWGN:
\begin{align}
x_{\text{sig1}}(t) = &\big(0.01 + 0.04\frac{t}{T} \big)\cos(2\pi 180 t) + \nonumber\\
&0.3 \cos\left(2\pi \big(\frac{50t}{T} + 300\big) t\right) + \mathcal{N}(0,0.3).
\label{eq:formula_BC_sig}
\end{align}
\noindent \Cref{fig:example_BC} shows the signal and the corresponding components, and \cref{fig:example_BC_decomp} shows the decompositions by different methods (zoomed in on the time axis for better visualization). \OUR~is able to accurately extract both the narrowband and broadband component, while VMD, SVMD and SSD are unable to achieve this without filtering out the noise, and EWT splits the broadband component in two.

\begin{figure}
\centering
    \centering
    \includegraphics[width=\linewidth]{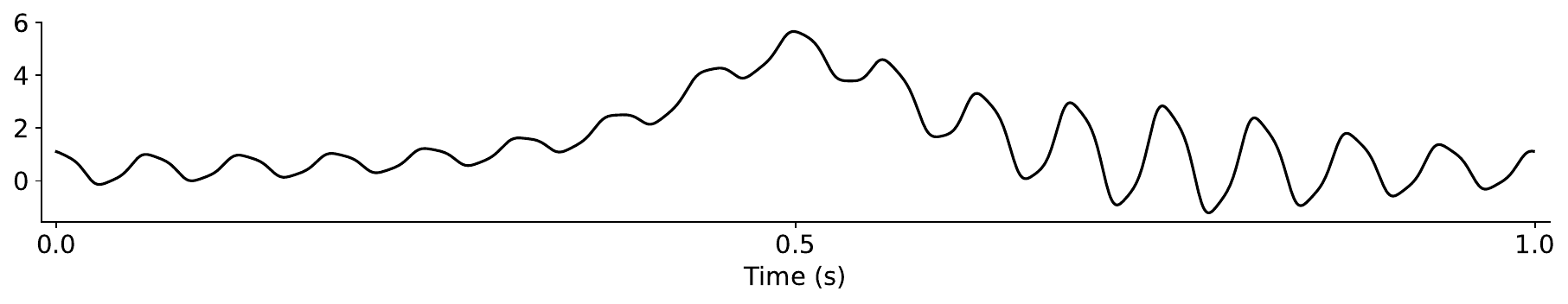}
    \caption{Non-stationary signal $x_{\text{sig2}}(t)$ with inter-wave frequency modulation, taken from \cite{Dragomiretskiy2014VariationalDecomposition}.}
    \label{fig:type_f0_K2_signal}
\end{figure}

\begin{figure*}
    \centering
    \includegraphics[width=\linewidth]{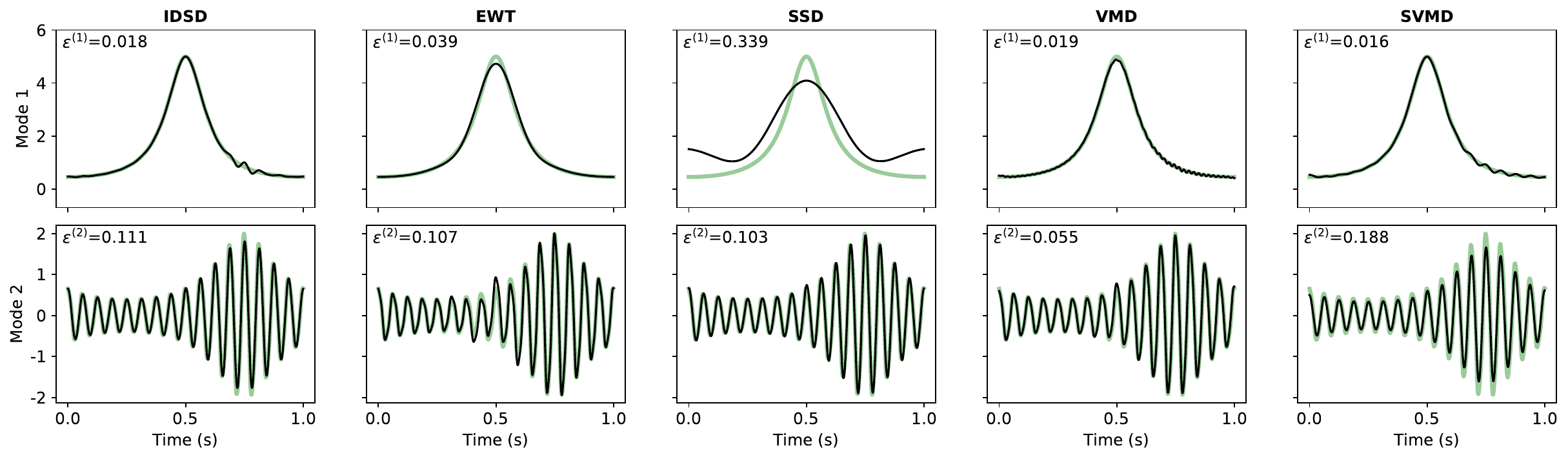}
\caption{Decomposition of signal $x_{\text{sig2}}(t)$ in two components by different methods (in black). Green denotes the ground-truth.}
\label{fig:casef1_k2_vmd}
\end{figure*}

 Next we take a signal that was presented by the authors of VMD~\cite[eq. 36]{Dragomiretskiy2014VariationalDecomposition}, being of the form:
\begin{equation}
x_{\text{sig2}}(t) = \frac{1}{1.2 + \cos\big(2\pi t)} + \frac{\cos(32\pi t + 0.2 \cos(64\pi t)\big)}{1.5 + \sin(2\pi t)},
\label{eq:formula_f_k2}
\end{equation}
containing a low-frequency component and an AM-FM component, see \cref{fig:type_f0_K2_signal}.
\Cref{fig:casef1_k2_vmd} shows that \OUR's performance is on par with EWT, VMD and SVMD for this signal. However, for VMD and SVMD, the optimal $\alpha$, resp., $\alpha_{\text{max}}$ had to be selected to achieve this performance, and for EWT and VMD, the number of components $K$ needed to be known in advance.
\subsection{Additional analyses}

We test several design choices of \OUR. First, we test the architecture and the type of output of the neural network that is part of \OUR. Second, we test how the number of trainable parameters affects performance. Third, we investigate two aspects of the training strategy, and fourth, we analyze the convergence of the residual energy.

\subsubsection{Model design}




~\\
\OUR~has built-in structure in two ways. First, thanks to the convolutional operations in the U-Net architecture it possesses a neighborhood prior which assumes that nearby bins in the input (magnitude) spectrum share information. Second, as our U-Net predicts a mask, rather than a component itself, the spectra of the generated components are guaranteed to be a subset of the signal's spectrum. To show the effectiveness of these design choices, we compare our baseline model against two different setups.

First, instead of using a U-Net architecture for mask prediction, we use a multi-layer perceptron (MLP) to make this prediction. This is similar to the strategy used in Neural Mode Estimation (NME)~\cite{Sun2023NeuralEstimation}, where an MLP is used to predict a mask in the DCT domain. We test two variants with a different number of trainable parameters. The small variant consists of the following layers: $\operatorname{Linear}(128) \rightarrow \operatorname{ReLU} \rightarrow \operatorname{Linear}(128) \rightarrow \operatorname{ReLU} \rightarrow \operatorname{Linear}(\mathrm{F}) \rightarrow \operatorname{Sigmoid_{\tau}}$, resulting in a mask predictor model with 148,481 trainable parameters. We compare it with \OUR~ with $C=32$, which has 124,160 trainable parameters. We also test against a larger MLP variant with 395,009 trainable parameters with the following architecture: $\operatorname{Linear}(256) \rightarrow \operatorname{ReLU} \rightarrow \operatorname{Linear}(256) \rightarrow \operatorname{ReLU} \rightarrow \operatorname{Linear}(256) \rightarrow \operatorname{ReLU} \rightarrow \operatorname{Linear}(\mathrm{F}) \rightarrow \operatorname{Sigmoid_{\tau}}$. 

\Cref{tab:comparing_architectures} shows the performance of these different models on our two test sets. The MLP architecture performances are far from IDSD's performance that uses a U-Net, even though both tested architectures contain more trainable parameters than the \OUR~model we compared against.

Second, instead of predicting a mask in frequency domain, we let the U-Net architecture predict a component in time domain directly. The residual is then found by subtracting the predicted component in time domain from the input residual, which is now also in time domain. The model is trained again with the MSE loss applied only on the dominant component, which we now defined as the component with most energy in time domain. \Cref{tab:comparing_architectures} shows that this direct prediction of components in time domain, heavily deteriorated performance compared to mask prediction in the frequency domain.

\begin{table*}
\centering
\caption{Normalized error values $\epsilon$ (median $\pm$ 1 standard deviation across test signals) of \OUR~that predicts frequency-domain masks using a U-Net architecture, against two sizes of multi-layer perceptrons that predict frequency-domain masks, and the same U-Net architecture that predicts components directly in time domain.} 
\begin{tabular}{l|c|c|c}
\toprule
 &  \bf Trainable parameters & $\mathbf{K=2}$ \bf test set & $\mathbf{K=5}$ \bf test set\\
\hline
\textbf{\OUR}:~U-Net ($C=32$) predicts frequency mask & 124,160  & $\mathbf{0.187 \pm 0.299}$  & $\mathbf{0.331 \pm 0.180}$\\
MLP predicts frequency mask & 148,481 & $0.733\pm0.381$ & $0.845 \pm 0.076$ \\
MLP (large) predicts frequency mask & 395,009 & $0.554 \pm 0.430$ & $0.781 \pm 0.113$ \\
\hline
\textbf{\OUR}: U-Net ($C=128$) predicts frequency mask & 1,983,488 &  $\mathbf{0.173 \pm 0.273}$ & $\mathbf{0.252 \pm 0.174}$ \\
U-Net ($C=128$) predicts component in time domain & 1,983,488 &  $0.404 \pm 0.222$ & $0.830 \pm 0.110$ \\
\bottomrule
\end{tabular}
\label{tab:comparing_architectures}
\end{table*}

\subsubsection{Effect of model size}

\begin{figure}
    \centering
\includegraphics[width=\linewidth,trim={0cm, 0cm, 0cm, 0.cm},clip]{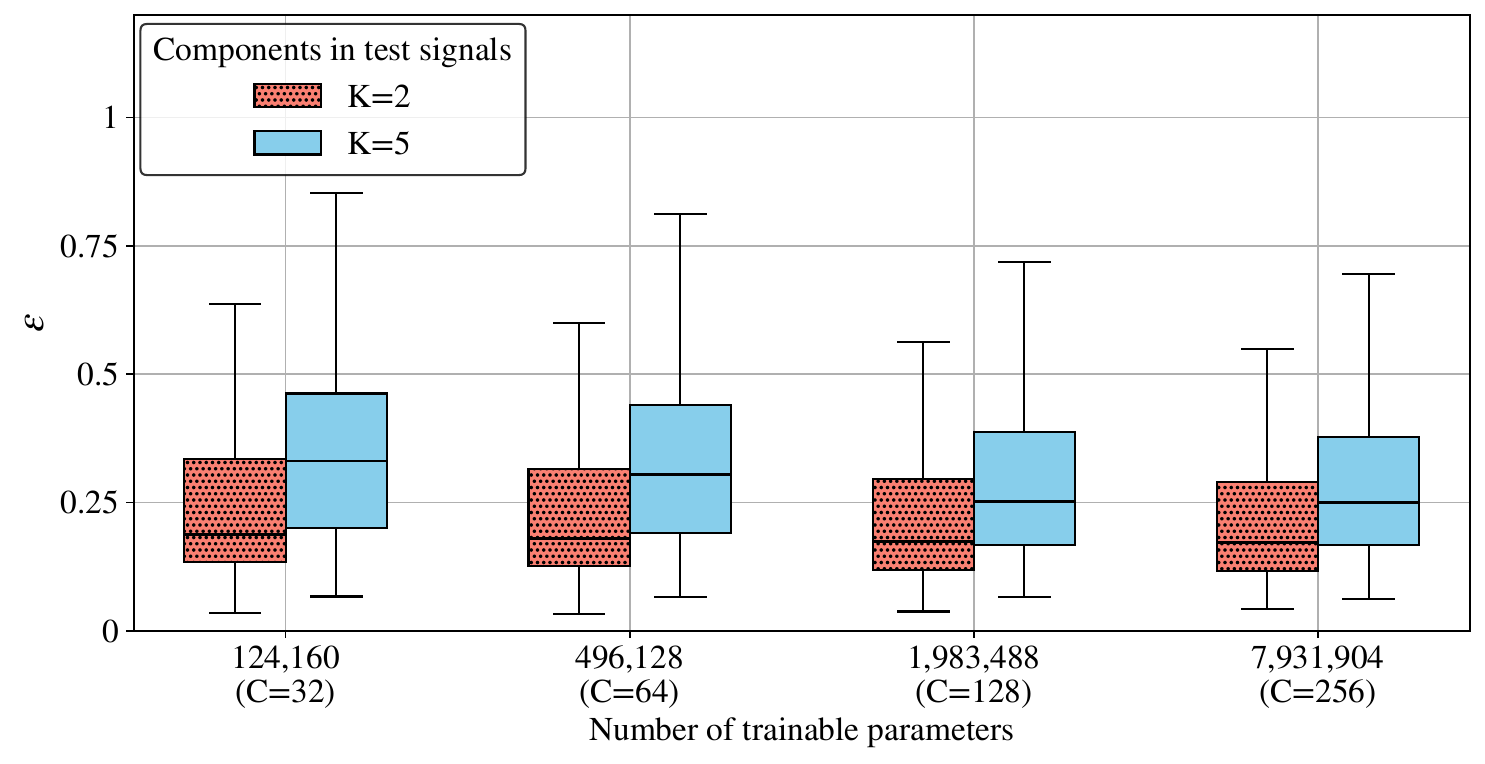}
    \caption{The relative error of \OUR~as a function of the trainable parameters in the U-Net architecture, which are varied by changing the number of channels $C$ in each convolutional layer.}
    \label{fig:error_per_C}
\end{figure}

The hyperparameters of the U-Net architecture were introduced in \cref{sec:architecture}, being  $B$ (number of encoder/decoder blocks), $L$ (kernel size of the convolutions), $P$ (pooling factor), and $C$ (number of convolutional channels), see also \cref{fig:Unet}. Changing $C$, the number of convolutional channels affects the number of trainable parameters while keeping the receptive field constant. Therefore, to test the effect of the number of trainable parameters on model performance, we vary $C$ in $\{32, 64, 128, 256\}$, while keeping $(B,L,P)=(3,7,3)$, and report the relative error of the model on both test sets. \Cref{fig:error_per_C} shows that test set performance for the $K=2$ test set does not benefit much from adding more trainable parameters, while performance for the more difficult test set with $K=5$ components per signal shows a slight trend towards better performance when using more trainable parameters, but performance remains similar for $C=128$ and $C=256$.

\subsubsection{Training strategy}
\label{sec:ablation_multiple_losses}

We test two aspects of our training strategy. First, we investigate the effect of the number of components used  to generate training signals. Second, we compare two variants of applying the loss function during training of the neural network. 

For the first test, we generate training data on the fly either with maximally $K=5$ components (as done in our proposed model), or with a fixed $K=1$, $K=2$, or $K=5$ components, while testing these models on our two test sets with $K=2$ and $K=5$ components.
\Cref{tab:comparing_training_generator} shows the results of this experiment. It can be seen that models for which $K$ was matched in training and testing (i.e. training and testing on signals with exactly $K$ components) outperformed models for which $K$ was different between training and testing. However, using a model trained on signals with a varying number of components (up to 5) resulted in an error at least as low as the errors of both specialized models for each test set (i.e. matching $K$ during training and testing). This is a useful property, as it enables using a single model for extracting any number of $K$ components during deployment. In other words, it does not require the user to train a new model every time a new number of components $K$ is expected during deployment. 

Second, we test the effect of training our model using an MSE loss on the dominant component only, as opposed to using the model to sequentially predict all components and use an MSE loss on all components in the signal (as explained in \cref{sec:training_phase}). \Cref{tab:comparing_training_loss} shows that our proposed model (with a loss only on the dominant component) slightly outperforms the model when trained on all components. So apart from having better performance when using only the dominant component in the loss, the training phase is also faster using this loss, as opposed to using \OUR~sequentially during training to extract all components for loss computation. 

\begin{table*}
\centering
\caption{Comparison of different strategies during data generation for training the model. The model that trains with signals that vary in the number of components (between 1 and maximally 5), performs best on both test sets.}
\begin{tabular}{l|c|c}
\toprule
 & $\mathbf{K=2}$ \bf test set & $\mathbf{K=5}$ \bf test set\\
\hline
\OUR~($C=128$): train with max. $K=5$ components & $\mathbf{0.173 \pm 0.273}$ & $\mathbf{0.252 \pm 0.174}$  \\
Train with $K=1$ component & $0.184 \pm 0.280$ & $0.420 \pm 0.185$ \\
Train with $K=2$ components & $0.175 \pm 0.264$ & $0.284 \pm 0.196$  \\
Train with $K=5$ components & $0.261 \pm 0.463$ & $0.256 \pm 0.179$  \\
\bottomrule
\end{tabular}
\label{tab:comparing_training_generator}
\end{table*}

\begin{table*}
\centering
\caption{\OUR~is trained with an MSE loss only on the dominant component. This results in faster and more stable training than using a loss on all components, which achieves similar performance.}
\begin{tabular}{l|c|c}
\toprule
 & $\mathbf{K=2}$ \bf test set & $\mathbf{K=5}$ \bf test set\\
\hline
\OUR~($C=128$): train with loss on dominant component & $\mathbf{0.173 \pm 0.272}$ & $\mathbf{0.252 \pm 0.174}$  \\
Train with loss on all components & $0.174 \pm 0.269$ & $0.264 \pm 0.170$ \\
\bottomrule
\end{tabular}
\label{tab:comparing_training_loss}
\end{table*}

\subsubsection{Convergence of residual energy}

In \cref{sec:stopping} we theoretically showed that our sequential mode extraction guarantees a reduction (or no change) in residual energy at every step that extracts an additional component. To verify this theoretical proof experimentally, we sequentially apply \OUR~8 times on each signal in the test set of $K=5$ components, and plot the residual energy at each step. Note that we, on purpose, apply the model more times than the ground-truth number of components, as there is no guarantee that the model will perfectly extract the $K=5$ components in exactly 5 steps. So applying \OUR~more than 5 times enables us to see whether or not the model is in general able to extract all components within the expected number of steps.

\begin{figure}
    \centering
\includegraphics[width=0.8\linewidth,trim={0cm, 0cm, 0cm, 0.cm},clip]{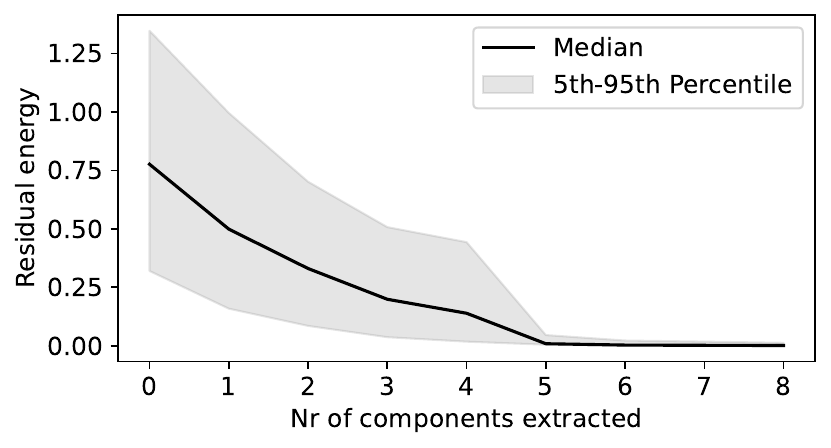}
    \caption{The residual energy $\frac{1}{N}\sum r^{(k)}[n]$ after extracting a certain number of components. We show the median taken over the 10,000 test signals with $K=5$ ground-truth components. The residual energy is guaranteed to go down after extracting an additional component, as shown in \cref{sec:stopping}, which is experimentally verified here.}
    \label{fig:residual_energy}
\end{figure}

\Cref{fig:residual_energy} shows the median of the residual energy over the 10,000 test set signals, and the $5-95$ percentile. It can be observed that the residual energy indeed decreases over steps. Also, after extracting $5$ components, the residual energy has become negligible in the majority of the signals (seen from the very small percentile range), suggesting that the model indeed extracted most energy within the expected number of 5 steps. The remaining small amount of energy can still be explained by the additive noise present in the signals.

\section{Experiments on real data}

\begin{figure*}
    \centering
    \includegraphics[width=\linewidth,trim={0cm 1cm 0cm 0cm},clip]{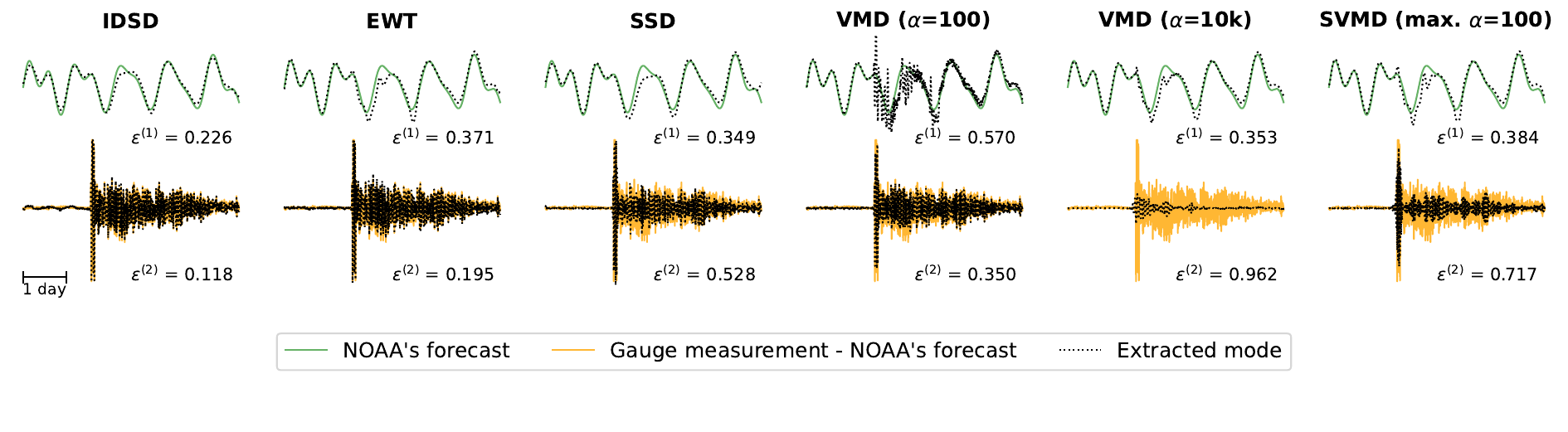}
    \caption{Decomposition of the tidal gauge measurement in two components by different methods (in black). First row: NOAA's forecasting of the tidal wave in green. Second row: the gauge measurement minus the forecasting in orange. IDSD extracts two components that closely follow the tidal forecasting and the tsunami component.}
    \label{fig:all_tsunami_res}
\end{figure*}

\subsection{Tidal waves}
\label{sec:tidal_data}

We analyze \OUR~on data measured with a tidal gauge from the National Oceanic and Atmospheric Administration (NOAA), placed in the Kahului Harbor in Maui, Hawaï\footnote{Data were acquired from \url{https://tidesandcurrents.noaa.gov/waterlevels.html?id=1615680}.}.
On 11 March 2011, a tsunami took place, which was visible on the measurement as an additive component on top of the tidal waves. We took the data starting from 10 March up to and including 14 March as test data. Data were sampled at 6 minutes intervals.
We treat NOAA's forecasting as the tidal component, and the full measurement minus the forecasting as the additive tsunami component. 

We used our three (varying in randomized seed) models (with $C=128$) that were trained on the synthetic data as a base. To further align \OUR~to the test distribution $p_{\mathrm{test}}$ from which the tidal gauge data originates, we finetuned ($\text{lr}_{\text{start}}=\mbox{5e-4}$) \OUR~on batches of 1024 5-day windows sampled randomly from measurements between 9 February 2011 and 9 March 2011 (so these data do not overlap with the data we use for testing, and they only contain regular tidal waves). We added AWGN $\mathcal{N}(0,1)$ to the input and trained until the MSE with NOAA's forecasted signal (i.e. the tidal component) plateaud. The fact that \OUR~only requires the ground-truth of only the dominant component in its loss function now appears useful, as a tsunami component is unavailable between 9 February and 9 March 2011. 

\Cref{fig:all_tsunami_res} shows that \OUR~can extract both the narrowband tidal component and the broadband tsunami component with the lowest average error (seed 1: $\epsilon\!=\!0.180$, seed 2: $\epsilon\!=\!0.172$  (visualized) and seed 3: $\epsilon\!=\!0.171$). For VMD we observe the same behavior as was seen from the synthetic data experiment in \cref{fig:example_BC_decomp}; VMD can either reasonably well extract the broadband tsunami component (for $\alpha=100$), or the narrowband tidal component (for $\alpha=10k$), but no other values of $\alpha$ improved extraction for both components simultaneously. EWT is the best-performing method among the baselines, with an average error of $\epsilon\!=\!0.283$. However, its decomposition relies on choosing the correct setting for $K$. When rerunning EWT while setting $K$ at different values, the average error changed to $\epsilon\!=\!0.437$, $\epsilon\!=\!0.542$, and $\epsilon\!=\!0.579$, for $K\!=\!3$, $K\!=\!4$, and $K\!=\!5$, respectively, which shows EWT's sensitivity to choosing the correct number of components.

We are interested to see how well \OUR~can decompose this tidal gauge measurement, without finetuning at all, i.e. directly deploying the model that is only trained on the synthetic signals. The relative errors increased from $(\epsilon^{(1)},\epsilon^{(2)})=(0.226,0.118)$ to $(\epsilon^{(1)},\epsilon^{(2)})=(0.323,0.169)$ when we omitted the finetuning step. This shows us two things. First, without finetuning \OUR~is still able to better extract the tidal and tsunami component from the measurement than all baselines, showing the strength of \OUR~on real data when only having synthetic data available for training. Second, finetuning on real data did improve performance, showing IDSD's ability to learn data-driven component priors. 

\begin{figure*}[h]
\centering
\includegraphics[width=\linewidth]{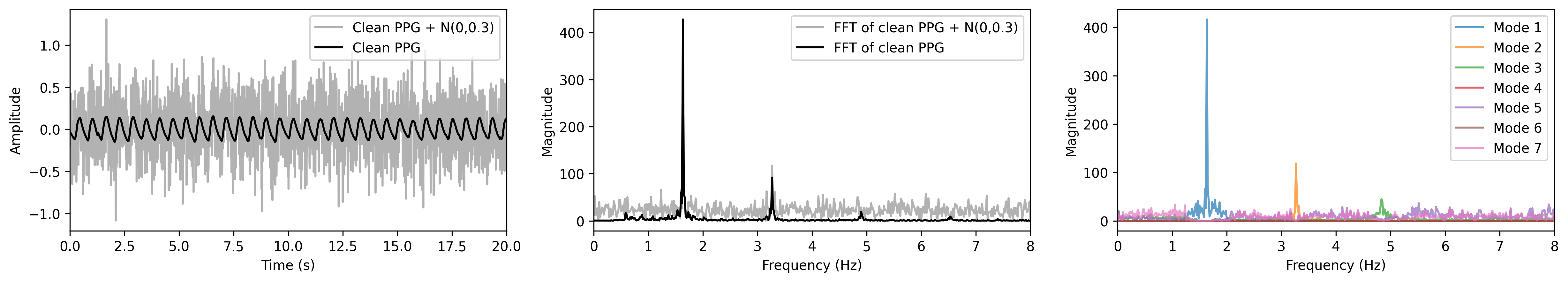}
\caption{Left: example of a finger PPG trace (black) with AWGN ($\mathcal{N}(0,0.3)$; grey). Middle: Corresponding frequency spectra. Right: Spectrum of the noisy PPG signal, decomposed by IDSD. The IDSD components clearly extract the fundamental heart rate (at 1.6 Hz; 96 beats per minute) and its harmonics.}
\label{fig:ppg_idsd}
\end{figure*}

\begin{figure*}[h]
    \centering
    \includegraphics[width=\linewidth,trim={0cm 0cm 0cm 0cm},clip]{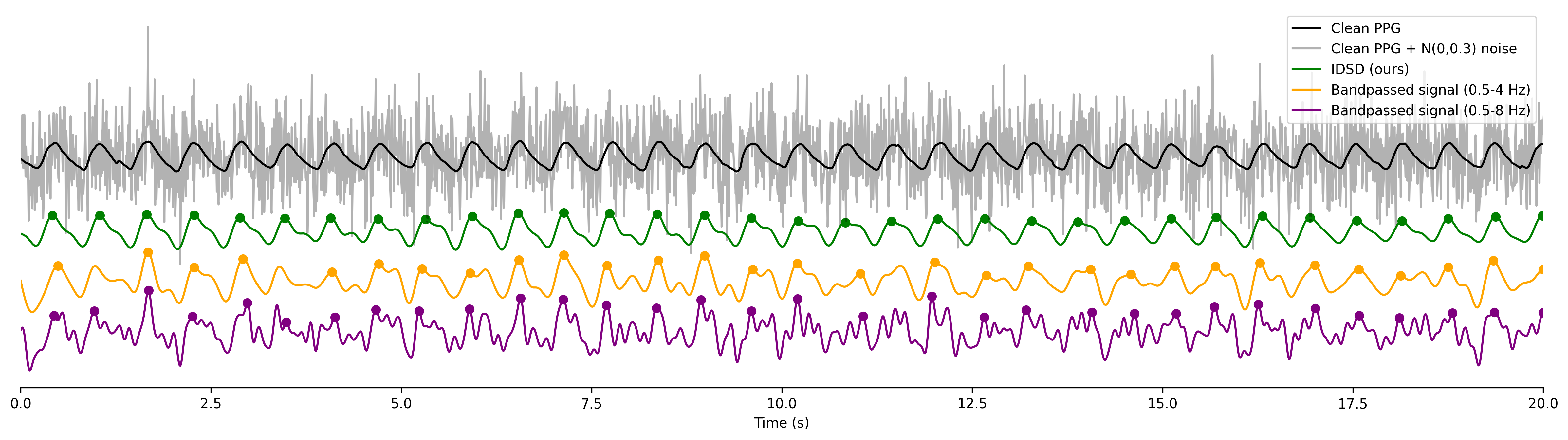}
    \caption{Clean and noisy PPG, and the denoised results from applying IDSD and bandpass filtering on the noisy PPG. Detected peaks are indicated, and for reference the ECG lead with detected peaks is shown as well. The IDSD result much better retains the waveform shape of the clean PPG signal.}
    \label{fig:ppg_baselines}
\end{figure*}

\subsection{Physiological data}

We use recordings from the BIDMC dataset \cite{pimentel2017toward}, which is a randomly-selected subset of a large cohort of critically-ill patients in the MIMIC-II dataset \cite{saeed2011multiparameter}. The BIDMC set contains 8 minutes of simultaneous multi-sensor recordings of $n=53$ subjects (age range:~$19$–$90+$, 32 females). 

We use the first minute of finger photoplethysmography (PPG) and electrocardiography (ECG) of each subject, which are both sampled at 125~Hz. To remove drift and respiratory-related components, we applied a 4th order high-pass Butterworth filter with a cutoff frequency of 0.5~Hz using zero-phase filtering. 

We use ECG data (lead II) as a reference for R-peak timing. To detect these peaks, we use the neurokit peak detector~\cite{makowski2021neurokit2}. To detect peaks in the PPG signal, we use scipy's findpeaks algorithm (with min. distance set to 0.5~s).

The PPG data in the BIDMC dataset has a high signal-to-noise ratio and little artifacts as it is recorded in a hospital setting. Yet, PPG sensors are nowadays often used in wearable devices, in which noise and movement artifacts are much more apparent~\cite{charlton2022wearable}. To test IDSD's ability to extract physiologically-relevant information from such noisy PPG data, we added AWGN $\mathcal{N}(0,\sigma)$ to each PPG recording, with varying $\sigma$ levels: $\{0.05,0.1,0.2,0.3,0.4,0.5\}$. We refer to the original PPG recording without AWGN as the `clean PPG' signal.

A PPG signal that cleanly reflects cardiac activity is a periodic signal (at the heart rate frequency), including harmonics that spur the PPG waveform shape. We applied \OUR~on the \textit{noisy PPG} signals, and summed IDSD`s first extracted component, and all IDSD components with a peak at harmonic frequencies of the first component. We compare against 4th order Butterworth bandpass filters with pass-bands of $0.5-4$~Hz and $0.5-8$~Hz, being typical approaches for PPG denoising and pre-processing~\cite{luo2021ippg,abdulsadig2024novel,quamer2025multimodal}. 

\Cref{fig:ppg_idsd}-left shows 20~s of one of the PPG recordings (left), including the noisy-PPG (with $\sigma\!=\!0.3$; in grey), which was the input for \OUR. The related Fourier spectra are shown in \cref{fig:ppg_idsd}-middle, and the decomposed spectrum by \OUR~is shown in \cref{fig:ppg_idsd}-right. \OUR~identified the fundamental peak and two of its harmonics from the noise floor. \Cref{fig:ppg_baselines} shows the resulting denoised signal for \OUR~and the band-pass filters. We visually observe that IDSD shows more consistency across cardiac cycles than bandpass filtering, and the waveform shape better resembles the waveform of the clean PPG. In some cardiac cycles (e.g. between 7.5-10 seconds), the IDSD output shows a deflection point in the downward slope. This deflection is common in PPG measurements, and it is a combination of the dicrotic notch (associated with aortic valve closure), and the diastolic peak (associated with reflected pressure waves from the peripheral arterial system~\cite{Elgendi2012OnSignals}. Interestingly, in the clean PPG signal, these deflections were less visible, possibly caused by the low -- but still-present -- amount of noise in the clean PPG data.

\begin{figure}[h]
    \centering
    \includegraphics[width=0.9\linewidth,trim={0cm 0cm 0cm 0cm},clip]{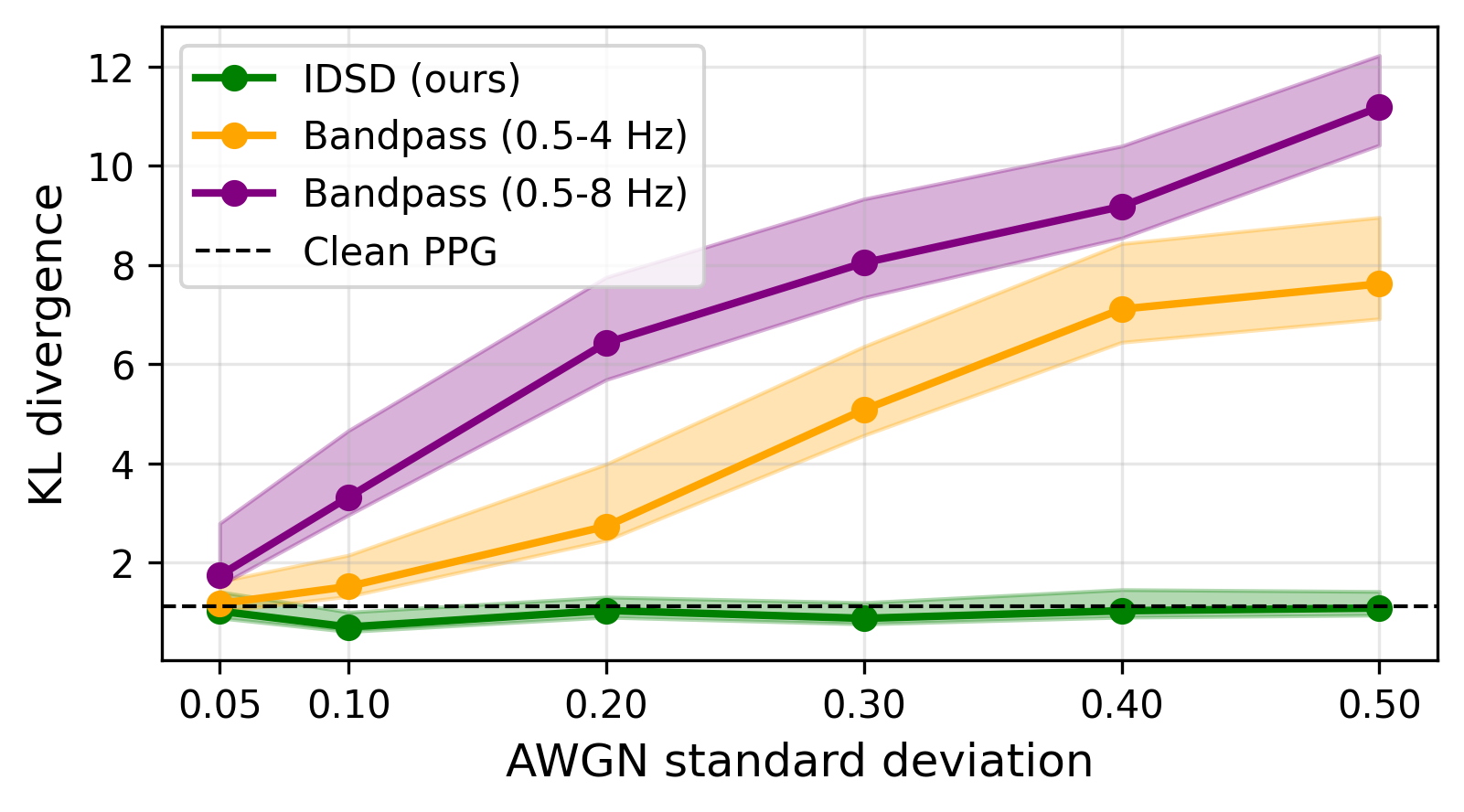}
    \caption{KL divergence (lower is better) between RR-interval times distribution from ECG and peak-interval times distribution after peak detection on IDSD and bandpass-filtered signals. Median and standard errors (across subjects) are shown. The peak-interval distribution from IDSD best resemble the RR-peak interval distributions from ECG.}
    \label{fig:KL_div}
\end{figure}

The detected peaks (indicated in \cref{fig:ppg_baselines} by dots) show timing differences between the peaks detected from IDSD's output and the bandpass filtered signals. The time between two R-peaks in ECG is called the RR-interval, and variability in the RR-interval (i.e. heart-rate variability (HRV)) is an important health parameter, as deviations in HRV have been associated with a range of physical and psychological conditions~\cite{shaffer2017overview}. As such, a good PPG peak detector should find a similar distribution of RR-interval times from PPG data, as found from peak detection in simultaneously-recorded ECG. We, therefore, compare the RR-interval times distribution from ECG to interval-time distributions extracted from IDSD's output and the bandpass filtered signals. For a quantitative comparison, we leverage the Kullback-Leibler (KL) divergence, which is zero for two perfectly-matching distributions, and larger than zero for non-matching distributions, with a higher KL-divergence indicating a worse match. \Cref{fig:KL_div} shows the median KL-divergence across subjects (incl. standard errors) across various noise levels, for \OUR and bandpass filtering. The dashed line indicates the KL-divergence between ECG RR-interval times and peak-interval times from the clean PPG (i.e. the original PPG signal without AWGN). While band-pass filtering with a tight band ($0.5-4$~Hz) performs as well as peak-detection on the clean PPG signal for very low noise levels, its performance heavily deteriorates once noise levels increase. On the contrary, the peak-time intervals from the \OUR signal are unaffected by the noise level, and result in the same match to the ECG peaks as using the clean PPG signal.

\section{Discussion}

In this work we propose \OUR:~Iterative Deep-learning-Based Signal Decomposition, a method for decomposing a univariate signal into its underlying components. We now discuss its properties and future directions.

\OUR~is a DL-based method, which results in a conceptually different algorithm than non-DL-based canonical methods like EMD, EWT, VMD and SSD. The DL-nature of \OUR~ensures that it can learn data-driven priors, as opposed to using fixed (e.g. narrowbanded) priors, for the components to be extracted. Our experiments on synthetic data and a tidal gauge measurement during a tsunami showed \OUR's~ability to decompose signals into various types of components. Moreover, we showed both on tidal gauge measurements and physiological measurements that \OUR~outperformed baselines, even when \OUR~was only trained on synthetic data. This shows the utility of \OUR, even in the absence of ground-truth components in a certain domain. Yet, in the presence of one ground-truth component, we showed on the tidal gauge measurement that \OUR~could further be improved by finetuning on it (the ground-truth component here was the tidal wave measurement taken from a different moment in time than the test data). The fact that finetuning further improved performance, shows IDSD's ability to learn data-driven priors.

The built-in structure in \OUR~ensures that the spectral support of the extracted components remains constrained by the spectrum of the input signal. This is conceptually similar to the working principle of EWT, where the Fourier domain is, as well, adaptively partitioned. This constrained is desirable in our context, as DL models are otherwise known to be able to generate spurious structures. In \cref{tab:comparing_architectures} we, indeed, showed that the incorporated structure both via mask prediction in frequency domain, and the U-Net architecture (as opposed to an MLP) contributed to the high performance. The fact that \OUR~employs the complete Fourier dictionary, which forms a complete basis for discrete-sampled signals, makes \OUR~sensor- and application-agnostic and, therefore, a versatile approach.

After training IDSD once, it can be deployed to extract a flexible and data-driven number of $K$ components, as \OUR~is applied multiple times until a stopping criterion is met. This in contrast to other DL-based methods NME~\cite{Sun2023NeuralEstimation} and $\operatorname{IRCNN}^{(+)}$~\cite{Zhou2024IRCNN:Network,Zhou2025IRCNN:Decomposition}, which require training a new model for every $K$. Other advantageous consequences of IDSD's setup is that the number of components to be extracted during deployment does not need to be known during training, and that the model size (i.e. the number of trainable parameters) is independent of $K$, which is not the case of IRCNN$^{+}$ and NME~\cite{Sun2023NeuralEstimation,Zhou2024IRCNN:Network,Zhou2025IRCNN:Decomposition}. Moreover, the sequential/greedy application of \OUR~during deployment provides the advantage that the first $k$ components do not differ dependent on how many total number of components $K$ are being extracted. Some other algorithms like EMD and SSD also possess this property. However, for both EWT and VMD, the extracted components depend on which $K$ was chosen (as was shown for EWT in \cref{sec:tidal_data}). Yet, non-greedy extraction in EWT and VMD, can bring advantages that are typical for non-greedy methods. In future research it could be investigated whether and how non-greedy component extraction could be embedded in \OUR~without giving up the property of being able to using a single \OUR~model for extracting a different number of components.

Thanks to the convolutional nature of \OUR, it can decompose signals that are of a different length than the ones used during training the model. Yet, if the test signal's length deviates a lot from the number of time samples used during training, the resolution of the spectral representation will also deviate considerably (i.e. it will be higher for longer signals). If the spectral resolution is much higher during testing than during training, the model might predicting very narrow band-pass masks. One way to circumvent this is to train \OUR~on synthetic data that comes close to the expected number of samples in the signals used during deployment. Additionally, one could also use Welch's estimator during deployment to create a smoother estimate of the spectral representation of the signal, or set a higher temperature value $\tau$ than the default $\tau=0.5$ during deployment. 

The speech field has also seen advances in DL-based methods for single-channel speech separation~\cite{Wang2018SupervisedOverview,Yang2025MultiscaleSeparation}. Even though proposed methods are typically dedicated to the application of speech separation in which often a fixed number of two or three sources is assumed, advances from this field could also further improve \OUR. For example, in the speech domain it has been observed that time-frequency DL mask predicting approaches introduce phase errors whenever the spectra of two or more sources overlap (as is usually the case in speech)~\cite{Choi2019Phase-awareU-net}. This is caused by the fact that from the phase of the mixture, the phase of the individual sources are not analytically resolvable without further constraints. Yet, the quantitative and qualitative accuracy with which IDSD recovered ground-truth components in this work seems to suggest that this issue is marginal for the type of signals tested here. However, future work could further explore how proposed solutions for phase errors in the speech domain~\cite{Roux2019TheSeparation,Choi2019Phase-awareU-net} could further improve IDSD's performance.

To conclude, we introduced IDSD, a DL-based signal decomposition method that predicts data-driven masks in frequency domain to extract signal components from an input spectrum. One single-trained IDSD model extracts a data-driven number of components thanks to its iterative deployment and an adaptive stopping criterion. The resulting decomposition is exact, i.e. the sum of the extracted components equals the input signal. Experimentally, we showed IDSD's superior performance against a large variety of baselines on both synthetic and real data. Moreover, we investigated and discussed architectural and training design choices of \OUR, and proved IDSD's convergence theoretically and empirically. \OUR~is a versatile signal decomposition method with utility in applications in which the number of components to be extracted is unknown and/or signal components do not nicely follow the narrowband assumptions as made by most other methods.

\bibliography{references, extra_refs}

@book{stein2011fourier,
  title={Fourier analysis: an introduction},
  author={Stein, Elias M and Shakarchi, Rami},
  volume={1},
  year={2011},
  publisher={Princeton University Press}
}

@article{antoni2006blind,
  title={Blind separation of vibration components: Principles and demonstrations.},
  author={Antoni, J},
  journal={The Shock and Vibration Digest},
  volume={38},
  number={4},
  pages={366--367},
  year={2006},
  publisher={Sage Publications, Inc.}
}

@article{pimentel2017toward,
  title={Toward a robust estimation of respiratory rate from pulse oximeters},
  author={Pimentel, Marco AF and Johnson, Alistair EW and Charlton, Peter H and Birrenkott, Drew and Watkinson, Peter J and Tarassenko, Lionel and Clifton, David A},
  journal={IEEE Transactions on Biomedical Engineering},
  volume={64},
  number={8},
  pages={1914--1923},
  year={2017},
  publisher={IEEE}
}

@article{saeed2011multiparameter,
  title={Multiparameter Intelligent Monitoring in Intensive Care II (MIMIC-II): a public-access intensive care unit database},
  author={Saeed, Mohammed and Villarroel, Mauricio and Reisner, Andrew T and Clifford, Gari and Lehman, Li-Wei and Moody, George and Heldt, Thomas and Kyaw, Tin H and Moody, Benjamin and Mark, Roger G},
  journal={Critical care medicine},
  volume={39},
  number={5},
  pages={952},
  year={2011}
}

@article{luo2021ippg,
  title={An iPPG-based device for pervasive monitoring of multi-dimensional cardiovascular hemodynamics},
  author={Luo, Jingjing and Zhen, Junjie and Zhou, Peng and Chen, Wei and Guo, Yuzhu},
  journal={Sensors},
  volume={21},
  number={3},
  pages={872},
  year={2021},
  publisher={MDPI}
}

@article{abdulsadig2024novel,
  title={A novel computational signal processing framework towards multimodal vital signs extraction using neck-worn wearable devices},
  author={Abdulsadig, Rawan S and Rodriguez-Villegas, Esther},
  journal={Scientific Reports},
  volume={14},
  number={1},
  pages={22368},
  year={2024},
  publisher={Nature Publishing Group UK London}
}

@article{quamer2025multimodal,
  title={A multimodal physiological dataset for non-invasive blood glucose estimation},
  author={Quamer, Waris and Tseng, Mu-Ruei and Vyas, Kathan and Dave, Darpit and Villegas, Carolina and McKay, Siripoom and DeSalvo, Daniel J and Erranguntla, Madhav and Cote, Gerard and Gutierrez-Osuna, Ricardo},
  journal={Scientific Data},
  volume={12},
  number={1},
  pages={1822},
  year={2025},
  publisher={Nature Publishing Group UK London}
}

@article{makowski2021neurokit2,
  title={NeuroKit2: A Python toolbox for neurophysiological signal processing},
  author={Makowski, Dominique and Pham, Tam and Lau, Zen J and Brammer, Jan C and Lespinasse, Fran{\c{c}}ois and Pham, Hung and Sch{\"o}lzel, Christopher and Chen, SH Annabel},
  journal={Behavior research methods},
  volume={53},
  number={4},
  pages={1689--1696},
  year={2021},
  publisher={Springer}
}

@article{shaffer2017overview,
  title={An overview of heart rate variability metrics and norms},
  author={Shaffer, Fred and Ginsberg, Jay P},
  journal={Frontiers in public health},
  volume={5},
  pages={258},
  year={2017},
  publisher={Frontiers Media SA}
}

@article{charlton2022wearable,
  title={Wearable photoplethysmography for cardiovascular monitoring},
  author={Charlton, Peter H and Kyriacou, Panicos A and Mant, Jonathan and Marozas, Vaidotas and Chowienczyk, Phil and Alastruey, Jordi},
  journal={Proceedings of the IEEE},
  volume={110},
  number={3},
  pages={355--381},
  year={2022},
  publisher={IEEE}
}

@inproceedings{Kingma2015Adam:Optimization,
    title = {{Adam: A method for stochastic optimization}},
    year = {2015},
    booktitle = {iclr},
    author = {Kingma, Diederik P and Ba, Jimmy}
}

@article{Hou2013Data-drivenAnalysis,
    title = {{Data-driven time-frequency analysis}},
    year = {2013},
    journal = {Applied and Computational Harmonic Analysis},
    author = {Hou, Thomas Y. and Shi, Zuoqiang},
    number = {2},
    month = {9},
    pages = {284--308},
    volume = {35},
    doi = {10.1016/j.acha.2012.10.001},
    issn = {10635203},
    arxivId = {1202.5621}
}

@article{Zhou2025IRCNN:Decomposition,
    title = {{IRCNN+: An Enhanced Iterative Residual Convolutional Neural Network for Non-stationary Signal Decomposition}},
    year = {2025},
    journal = {Pattern Recognition Letters},
    author = {Zhou, Feng and Cicone, Antonio and Zhou, Haomin},
    month = {9},
    pages = {328v-336},
    volume = {196},
    url = {http://arxiv.org/abs/2309.04782},
    arxivId = {2309.04782}
}

@article{Gilles2013EmpiricalTransform,
    title = {{Empirical wavelet transform}},
    year = {2013},
    journal = {IEEE transactions on signal processing},
    author = {Gilles, Jerome},
    number = {16},
    pages = {3999--4010},
    volume = {61},
    url = {http://arxiv.org/abs/2410.23534 http://dx.doi.org/10.1109/TSP.2013.2265222},
    doi = {10.1109/TSP.2013.2265222},
    arxivId = {2410.23534}
}

@article{Starck2005ImageApproach,
    title = {{Image decomposition via the combination of sparse representations and a variational approach}},
    year = {2005},
    journal = {IEEE Transactions on Image Processing},
    author = {Starck, Jean Luc and Elad, Michael and Donoho, David L.},
    number = {10},
    month = {10},
    pages = {1570--1582},
    volume = {14},
    url = {https://ieeexplore.ieee.org/abstract/document/1510691},
    doi = {10.1109/TIP.2005.852206},
    issn = {10577149},
    pmid = {16238062}
}

@article{Zhou2024IRCNN:Network,
    title = {{IRCNN: A novel signal decomposition approach based on iterative residue convolutional neural network}},
    year = {2024},
    journal = {Pattern Recognition},
    author = {Zhou, Feng and Cicone, Antonio and Zhou, Haomin},
    month = {11},
    pages = {110670},
    volume = {155},
    publisher = {Elsevier Ltd},
    doi = {10.1016/j.patcog.2024.110670},
    issn = {00313203}
}

@article{Yang2025MultiscaleSeparation,
    title = {{Multiscale Convolutional Fusion Network for Efficient Monaural Speech Separation}},
    year = {2025},
    journal = {IEEE Access},
    author = {Yang, Rui and Pan, Shanliang},
    pages = {127045--127053},
    volume = {13},
    publisher = {Institute of Electrical and Electronics Engineers Inc.},
    url = {https://ieeexplore.ieee.org/abstract/document/11072705},
    doi = {10.1109/ACCESS.2025.3587085},
    issn = {21693536}
}

@inproceedings{Sun2023NeuralEstimation,
    title = {{Neural Mode Estimation}},
    year = {2023},
    booktitle = {ICASSP, IEEE International Conference on Acoustics, Speech and Signal Processing - Proceedings},
    author = {Sun, Peng and Wen, Zhenyu and Zhou, Yejian and Hong, Zhen and Lin, Tao},
    volume = {2023-June},
    publisher = {Institute of Electrical and Electronics Engineers Inc.},
    isbn = {9781728163277},
    doi = {10.1109/ICASSP49357.2023.10094930},
    issn = {15206149}
}

@article{Elgendi2012OnSignals,
    title = {{On the Analysis of Fingertip Photoplethysmogram Signals}},
    year = {2012},
    journal = {Current Cardiology Reviews},
    author = {Elgendi, Mohamed},
    pages = {14--25},
    volume = {8}
}

@inproceedings{Choi2019Phase-awareU-net,
    title = {{Phase-aware speech enhancement with deep complex u-net}},
    year = {2019},
    booktitle = {ICLR},
    author = {Choi, Hyeong-Seok and Kim, Jang-Hyun and Huh, Jaesung and Kim, Adrian and Ha, Jung-Woo and Lee, Kyogu},
    url = {http://kkp15.github.io/DeepComplexUnet}
}

@inproceedings{Behboodi2020ReceptiveU-Net,
    title = {{Receptive field size as a key design parameter for ultrasound image segmentation with U-Net}},
    year = {2020},
    booktitle = {EMBC},
    author = {Behboodi, Bahareh and Fortin, Maryse and Belasso, Clyde J. and Brooks, Rupert and Rivaz, Hassan},
    pages = {2117--2120},
    publisher = {IEEE},
    isbn = {9781728119908}
}

@article{Bonizzi2014SingularDecomposition,
    title = {{Singular Spectrum Decomposition: A new method for time series decomposition}},
    year = {2014},
    journal = {Advances in Adaptive Data Analysis},
    author = {Bonizzi, Pietro and Karel, Joël M. H. and Meste, Olivier and Peeters, Ralf L. M.},
    number = {4},
    pages = {1450011},
    volume = {6},
    publisher = {World Scientific Pub Co Pte Lt},
    doi = {10.1142/s1793536914500113},
    issn = {1793-5369}
}

@techreport{LeRoux2015SparseDone,
    title = {{Sparse NMF-half-baked or well done?}},
    year = {2015},
    author = {Le Roux, Jonathan and Weninger, Felix and Hershey, John R},
    pages = {13--15},
    institution = {Mitsubishi Electric Research Labs (MERL), Cambridge, MA, USA, Tech. Rep., no. TR2015-023}
}

@article{Richardson2024SRMD:Decomposition,
    title = {{SRMD: Sparse Random Mode Decomposition}},
    year = {2024},
    journal = {Communications on Applied Mathematics and Computation},
    author = {Richardson, Nicholas and Schaeffer, Hayden and Tran, Giang},
    number = {2},
    month = {6},
    pages = {879--906},
    volume = {6},
    publisher = {Springer Nature},
    doi = {10.1007/s42967-023-00273-x},
    issn = {26618893},
    arxivId = {2204.06108}
}

@article{Nazari2020SuccessiveDecomposition,
    title = {{Successive variational mode decomposition}},
    year = {2020},
    journal = {Signal Processing},
    author = {Nazari, Mojtaba and Sakhaei, Sayed Mahmoud},
    month = {9},
    pages = {107610},
    volume = {174},
    publisher = {Elsevier B.V.},
    doi = {10.1016/j.sigpro.2020.107610},
    issn = {01651684}
}

@article{Wang2018SupervisedOverview,
    title = {{Supervised speech separation based on deep learning: An overview}},
    year = {2018},
    journal = {IEEE/ACM Transactions on Audio Speech and Language Processing},
    author = {Wang, Deliang and Chen, Jitong},
    number = {10},
    month = {10},
    pages = {1702--1726},
    volume = {26},
    publisher = {Institute of Electrical and Electronics Engineers Inc.},
    doi = {10.1109/TASLP.2018.2842159},
    issn = {23299290},
    arxivId = {1708.07524}
}

@article{Huang1998TheAnalysis,
    title = {{The empirical mode decomposition and the Hilbert spectrum for nonlinear and non-stationary time series analysis}},
    year = {1998},
    journal = {Proceedings of the Royal Society of London. Series A: mathematical, physical and engineering sciences},
    author = {Huang, Norden E and Shen, Zheng and Long, Steven R and Wu, Manli C and Shih, Hing H and Zheng, Quanan and Yen, Nai-Chyuan and Tung, Chi Chao and Liu, Henry H},
    number = {1971},
    pages = {903--995},
    volume = {454},
    url = {https://royalsocietypublishing.org/}
}

@techreport{Roux2019TheSeparation,
    title = {{The Phasebook: Building Complex Masks via Discrete Representations for Source Separation}},
    year = {2019},
    author = {Roux, Le and Wichern, J ; and Watanabe, G ; and Sarroff, S ; and Hershey, A ; and Le Roux, Jonathan and Wichern, Gordon and Watanabe, Shinji and Sarroff, Andy and Hershey, John R},
    url = {http://www.merl.com}
}

@inproceedings{Ronneberger2015U-net:Segmentation,
    title = {{U-net: Convolutional networks for biomedical image segmentation}},
    year = {2015},
    booktitle = {MICCAI},
    author = {Ronneberger, Olaf and Fischer, Philipp and Brox, Thomas},
    pages = {234--241},
    volume = {9351},
    publisher = {Springer Verlag},
    isbn = {9783319245737},
    doi = {10.1007/978-3-319-24574-4{\_}28},
    issn = {16113349}
}

@article{Dragomiretskiy2014VariationalDecomposition,
    title = {{Variational mode decomposition}},
    year = {2014},
    journal = {IEEE Transactions on Signal Processing},
    author = {Dragomiretskiy, Konstantin and Zosso, Dominique},
    number = {3},
    pages = {531--544},
    volume = {62},
    doi = {10.1109/TSP.2013.2288675},
    issn = {1053587X}
}

@article{Le2018WhatNetworks,
    title = {{What are the Receptive, Effective Receptive, and Projective Fields of Neurons in Convolutional Neural Networks?}},
    year = {2018},
    author = {Le, Hung and Borji, Ali},
    month = {4},
    url = {http://arxiv.org/abs/1705.07049},
    arxivId = {1705.07049}
}
\bibliographystyle{ieeetr}

\newpage
\appendices

\section{Receptive field of U-Net}
\label{app:receptive_field}

\noindent The effective receptive field $RF_i$ at layer $i$ (counting from 1) of a convolutional neural network can be computed as~\cite{Le2018WhatNetworks}:
\begin{equation}
RF_i = RF_{i-1} + (Q_i-1) \prod_{j=1}^{i-1} S_j, \quad \text{with} \quad RF_0 = 1,
\label{eq:RF}
\end{equation}

\noindent $Q_i$ the kernel size of layer $i$, and $S_j$ the stride. Our U-Net encoder contains $3B+2$ layers ($B$ blocks of two convolutional layers plus a pooling layer, and a bottleneck block of two convolutional layers).
All convolutional layers have a kernel size of $L$, and a stride of 1, while pooling was implemented with max. pooling layers with a kernel size of $P$, and a stride of $P$. Formally, in our encoder:

\begin{equation}
    Q_i = \begin{cases} P & \text{if } i\%3=0 \\ 
    L & \text{otherwise} \end{cases}
\end{equation}

\begin{equation}
    S_i = \begin{cases} P & \text{if } i\%3=0 \\ 1 & \text{otherwise} \end{cases}
\end{equation}

We use \cref{eq:RF} to analyze the effective RF for our U-Net encoder for different settings of $L$, $P$ and $B$, see \cref{tab:RF_computations}. 

Since our synthetically-generated signals contain $1024$ samples (duration of 1~s, and a sampling frequency of $fs=1024$~Hz), the number of frequency bins $F$ in the positive side of the spectrum equals $F=1024/2+1=513$. We aim for an effective RF that is close to this value, as the effective RF field has been shown an important parameter for performance of U-Net models~\cite{Behboodi2020ReceptiveU-Net}.
From \cref{tab:RF_computations} it can be seen that different combinations of values for $B$, $P$ and $L$ can be candidates for this requirement, e.g. $(B,P,L)=(3,3,7)$, and $(B,P,L)=(4,2,9)$. Preliminary tests revealed marginal differences between these two settings, while the latter has more trainable parameters (caused by the higher number of blocks $B$, and larger kernel size $L$). We, therefore, selected $(B,P,L)=(3,3,7)$ in this work, indicated in bold in \cref{tab:RF_computations}.

\renewcommand{\arraystretch}{0.75} 

\begin{table}[ht]
\centering
\caption{Effective receptive field of the U-Net encoder for different combinations of $B$, $P$, and $L$.
The number of layers equals $3B+2$.}
\label{tab:RF_computations}
\scriptsize
\begin{tabular}{ccccc}
\toprule
$B$ & Number of layers & $P$ & $L$ & RF of encoder \\
\midrule
\multirow{12}{*}{1} & \multirow{12}{*}{5}
    & \multirow{4}{*}{2} & 3 &  14 \\
  & &                    & 5 &  26 \\
  & &                    & 7 &  38 \\
  & &                    & 9 &  50 \\
\cmidrule{3-5}
  & & \multirow{4}{*}{3} & 3 &  19 \\
  & &                    & 5 &  35 \\
  & &                    & 7 &  51 \\
  & &                    & 9 &  67 \\
\cmidrule{3-5}
  & & \multirow{4}{*}{4} & 3 &  24 \\
  & &                    & 5 &  44 \\
  & &                    & 7 &  64 \\
  & &                    & 9 &  84 \\
\midrule
\multirow{12}{*}{2} & \multirow{12}{*}{8}
    & \multirow{4}{*}{2} & 3 &   32 \\
  & &                    & 5 &   60 \\
  & &                    & 7 &   88 \\
  & &                    & 9 &  116 \\
\cmidrule{3-5}
  & & \multirow{4}{*}{3} & 3 &   61 \\
  & &                    & 5 &  113 \\
  & &                    & 7 &  165 \\
  & &                    & 9 &  217 \\
\cmidrule{3-5}
  & & \multirow{4}{*}{4} & 3 &  100 \\
  & &                    & 5 &  184 \\
  & &                    & 7 &  268 \\
  & &                    & 9 &  352 \\
\midrule
\multirow{12}{*}{3} & \multirow{12}{*}{11}
    & \multirow{4}{*}{2} & 3 &    68 \\
  & &                    & 5 &   128 \\
  & &                    & 7 &   188 \\
  & &                    & 9 &   248 \\
\cmidrule{3-5}
  & & \multirow{4}{*}{3} & 3 &   187 \\
  & &                    & 5 &   347 \\
  & &                    & 7 &   \bf 507 \\
  & &                    & 9 &   667 \\
\cmidrule{3-5}
  & & \multirow{4}{*}{4} & 3 &   404 \\
  & &                    & 5 &   744 \\
  & &                    & 7 & 1\,084 \\
  & &                    & 9 & 1\,424 \\
\midrule
\multirow{12}{*}{4} & \multirow{12}{*}{14}
    & \multirow{4}{*}{2} & 3 &   140 \\
  & &                    & 5 &   264 \\
  & &                    & 7 &   388 \\
  & &                    & 9 &   512 \\
\cmidrule{3-5}
  & & \multirow{4}{*}{3} & 3 &   565 \\
  & &                    & 5 & 1\,049 \\
  & &                    & 7 & 1\,533 \\
  & &                    & 9 & 2\,017 \\
\cmidrule{3-5}
  & & \multirow{4}{*}{4} & 3 & 1\,620 \\
  & &                    & 5 & 2\,984 \\
  & &                    & 7 & 4\,348 \\
  & &                    & 9 & 5\,712 \\
\bottomrule
\end{tabular}%
\end{table}

\section{Benchmarking $\text{IRCNN}^{+}$}
\label{app:benchmark_ircnn}
We converted the available Keras code for $\text{IRCNN}^{+}$~\cite{Zhou2025IRCNN:Decomposition} to Pytorch and benchmarked it by training and evaluating $\text{IRCNN}^{+}$ on the train, resp., validation set of their \textit{Dataset\_2}. The authors report an average MAE of 0.0338 (our implementation: 0.0304) and RMSE of 0.0627 (our implementation: 0.0439) when using a kernel size of 32. 
The better performance by our implementation could be thanks to the fact that we used a learning rate scheduler and trained until convergence instead of stopping at a fixed epoch. 
Next, we trained $\text{IRCNN}^{+}$ with varying sets of hyperparameters on our data generator, and report the best-performing model on our synthetic test dataset with $K=2$ components. Note that the Keras implementation of the $\text{IRCNN}^{+}$ model was only provided for extracting two components. As the architecture of $\text{IRCNN}^{+}$ depends on $K$, testing $\text{IRCNN}^{+}$ for $K=5$ would require a re-design of the model. As such $\text{IRCNN}^{+}$ was only tested for $K=2$ components.

\section{Performance on specific component types}
\label{app:subsets}

\begin{figure*}[ht]
    \centering
\includegraphics[width=\linewidth,trim={0cm, 0cm, 0cm, 0.cm},clip]{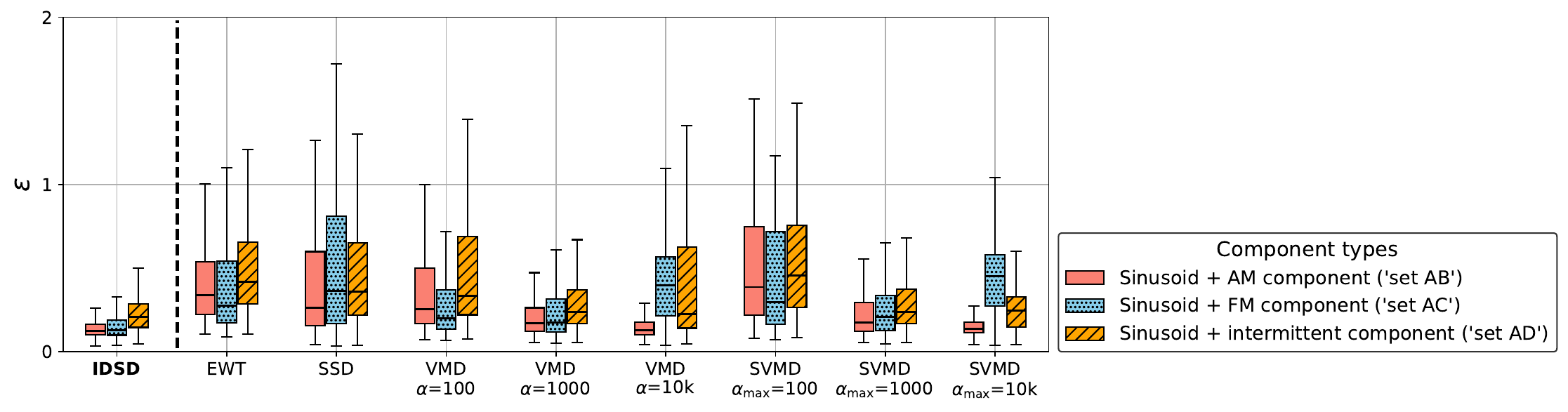}
    \caption{Relative error distributions of different test sets that each contain 10,000 signals, each with two different type of components and AWGN ($\sigma=0.1$).}
    \label{fig:subsets}
\end{figure*}

\Cref{fig:subsets} compares the relative errors of \OUR~(seed 1) to the best performing baselines from \cref{fig:synthetic_K2_K5} for sets AB, AC and AD. Set AB, which contains relatively narrowband components, is decomposed best by IDSD, and both VMD and SVMD with high values for $\alpha_{(\text{max})}$, while for lower values of $\alpha_{(\text{max})}$, these algorithms start to perform worse. Performance on sets AC and AD -- which contain FM components, resp., intermittent components -- on the other hand is lower for all baselines (i.e. larger errors), compared to \OUR.

\end{document}